\documentclass[prb,groupedaddress,preprint,floatfix]{revtex4}
\usepackage{graphicx}
\usepackage{color}
\usepackage{bm,amsmath,amssymb,enumitem,mathtools}

 \usepackage{xcolor}
\usepackage{multirow}

\usepackage{ulem} 

\begin{document}

\title{Towards a deeper fundamental understanding \\ of (Al,Sc)N ferroelectric nitrides}

\author{Peng Chen}
\affiliation{Smart Ferroic Materials Center, Institute for Nanoscience \& Engineering and Department of Physics, University of Arkansas, Fayetteville, Arkansas 72701, USA}
\affiliation{Department of Physics, Guangdong Technion -- Israel Institute of Technology, Shantou, Guangdong Province, China}
\author{Dawei Wang}
\affiliation{Smart Ferroic Materials Center, Institute for Nanoscience \& Engineering and Department of Physics, University of Arkansas, Fayetteville, Arkansas 72701, USA}
\affiliation{University of Houston, Houston, Texas 77204}
\author{Alejandro Mercado Tejerina}
\affiliation{Smart Ferroic Materials Center, Institute for Nanoscience \& Engineering and Department of Physics, University of Arkansas, Fayetteville, Arkansas 72701, USA}
\author{Keisuke Yazawa}
\affiliation{Department of Metallurgical and Materials Engineering, Colorado School of Mines, Golden, Colorado 80401, USA}
\affiliation{Materials Science Center, National Renewable Energy Laboratory, Golden, Colorado 80401, USA}
\author{Andriy Zakutayev}
\affiliation{Materials Science Center, National Renewable Energy Laboratory, Golden, Colorado 80401, USA}
\author{Charles Paillard}
\affiliation{Smart Ferroic Materials Center, Institute for Nanoscience \& Engineering and Department of Physics, University of Arkansas, Fayetteville, Arkansas 72701, USA}
\affiliation{Université Paris-Saclay, CentraleSupélec, CNRS, Laboratoire SPMS, 91190, Gif-sur-Yvette, France}
\author{L. Bellaiche}
\affiliation{Smart Ferroic Materials Center, Institute for Nanoscience \& Engineering and Department of Physics, University of Arkansas, Fayetteville, Arkansas 72701, USA}
\affiliation{Department of Materials Science and Engineering, Tel Aviv University, Ramat Aviv, Tel Aviv 6997801, Israel}

\begin{abstract}
Density Functional Theory (DFT) calculations, within the virtual crystal alloy approximation, are performed, along with the development of a Landau-type model employing a symmetry-allowed analytical expression of the internal energy and having parameters being determined from first principles, to investigate properties and energetics of Al$_{1-x}$Sc$_x$N ferroelectric nitrides in their hexagonal forms. These DFT computations and this model predict the existence of two different types of minima, namely the 4-fold-coordinated wurtzite (WZ) polar structure and a 5-times paraelectric hexagonal  phase (to be denoted as H5), for any  Sc composition up to 40\%. The H5 minimum progressively becomes the lowest energy state within hexagonal symmetry as the Sc concentration increases from 0 to 0.4. Furthermore, the model points out to several key findings. Examples include the crucial role of the coupling between polarization and strains to create the WZ minimum, in addition to polar and elastic energies, and that  the origin of the H5 state overcoming the  WZ phase as the global minimum within hexagonal symmetry when increasing the Sc composition mostly lies in the compositional dependency of only two parameters, one linked to the polarization and another one being purely elastic in nature. Other examples are that forcing Al$_{1-x}$Sc$_x$N systems  to have no or a weak change in lattice parameters when heating them allows to reproduce well their finite-temperature polar properties, and that a value of the axial ratio close to that of the ideal WZ structure does imply a large polarization at low temperatures but not  necessarily at high temperatures because of the ordered-disordered character of the temperature-induced formation of the WZ state. Such findings should allow for a better fundamental understanding of (Al,Sc)N ferroelectric nitrides, which may be put in use to design efficient devices having, e.g., low operating voltages.
\end{abstract}

 
\maketitle

\section{Introduction}

In 2002, a first-principles study was devoted to the search for a hexagonal metastable state of pure ScN, in addition to its well-known rocksalt (cubic) ground state \cite{Farrer}. 
The results were unexpected in the sense that the predicted hexagonal metastable state was neither the four-time-coordinated wurtzite (WZ) polar structure that typical III-V nitrides \cite{Nakamura,Edgar,Wright}, such as AlN, GaN and InN, are known to adopt with a $c/a$ axial ratio of $\simeq$  1.6 nor the 3-time-coordinated paraelectric layered structure that BN possesses with a huge $c/a$ of about 2.6 \cite{Ohba}. The simulations rather yielded a five-time-coordinated hexagonal phase (coined here as H5) phase that is paraelectric too but with a much smaller axial ratio  of about 1.2. Several other {\it ab-initio} simulations \cite{Ranjan,Ranjan2,Ranjan3,Daoust,Tasnadi,Noor,Zhijun,Zhijun2} then proposed that solid-solutions of ScN with typical III-V nitrides may induce, for some Sc concentrations, ferroelectricity -- that is, the reversibility of a spontaneous polarization under an applied electric field. Some of these works also suggested that such mixing may lead to other  technologically-useful properties and novel phenomena, including large piezoelectricity, elasto-optic coefficients and electro-optic conversion, as well as tunable electronic band-gap and even ultrahigh energy storage density.

In 2009, enhancement of piezoelectricity was indeed measured in disordered (Al,Sc)N systems when varying the Sc composition \cite{Akiyama,Akiyama2}, followed ten years later by the experimental demonstration that these nitride alloys do show ferroelectricity for some Sc concentrations \cite{Fichtner}. These discoveries have since led to a flurry of activities on these ferroelectric nitrides  and references therein), especially since not only these systems are compatible with Si technology but also because their spontaneous polarization is larger than 100 $\mu C/cm^2$, their polarization-{\it versus}-electric field hysteresis loops has a nearly square shape and they are thermally stable up to $\simeq$ 1400K  (see, e.g., Refs. \cite{Fichtner,Wolff,Islam,Guido,Zhu,Yazawa1,Yazawa2,Lee1,Lee2,Calderon}). These remarkable nitride semiconductors are therefore highly attractive for various applications, such as  micro/nano-actuators, devices based on the control of electrical polarization, as well as, for nonvolatile memory and harsh-environment technologies \cite{Drury,Drury2}.  

Despite all these works, a deep understanding of hexagonal ferroelectric nitrides is missing in our minds. For instance, one may wonder what are the mechanisms that govern the existence of two different hexagonal states with different coordinations, axial ratio and polarities within the same system? What interactions are responsible for the H5 state progressively overcoming the WZ state as the hexagonal phase of minimum energy when increasing Sc composition in (Al,Sc)N? Do ferroelectric nitrides differ from the better-known
ferroelectric perovskite oxides in terms of effects responsible for their polarization, but also for their finite-temperature properties? Are phenomenological models able to describe the hexagonal states of ferroelectric nitrides \cite{Kei-Andriy}? If yes, what are the degrees of freedom, their minimal orders and their interactions that are required for these phenomenological models to be accurate? 

The goal of this article is to provide an answer to all these questions, by (i) conducting first-principles calculations; (ii)  proposing a model based on an analytical expression of the internal energy that is valid for any hexagonal ferroelectric system ; (iii) determining the coefficients that appear in this expression in an {\it ab-initio way} for  (Al,Sc)N; (iv) taking advantage of this model by turning on and off some of its parameters; and (v) incorporating this model in Monte-Carlo simulations. As we will see, surprises are in store and require adopting a new thinking when investigating these nitride semiconductors.

This article is organized as follows. Section II describes the presently used first-principles methods. Notations to be employed, structures to be studied and the analytical expression of the aforementioned model are provided in Section III. Results of first-principles computations are given and discussed in Section IV, in addition to details explaining how to get the parameters of the model. The accuracy of this model is also demonstrated in Section IV, along with its proposed use to acquire a deep insight into ferroelectric nitrides and to compare them with perovsvkite oxides. Section V discusses other possible uses of this model, such as the determination of so-called hidden states and of finite-temperature properties. Finally, a summary is given in Section VI.

\section{Methods}

In this work, we performed density functional theory (DFT) calculations with the Quantum Espresso open-source software  \cite{Giannozzi2009}. For the exchange-correlation functional, we chose the Perdew-Burke-Ernzerhof (PBE) functional \cite{Perdew1996} and described the core electrons using Optimized Norm-Conserving Vanderbilt (ONCV) pseudopotentials\cite{Hamann2013}. We expanded the wavefunctions with a plane-wave basis up to a kinetic energy cutoff of 90 Ry. To integrate the Brillouin zone, we used a $10\times10\times8$ Monkhorst-Pack k-point mesh \cite{Monkhorst1976} and relaxed the atomic positions until the forces on each atom were below $1\times10^{-6}$ Hartree/Bohr. The mixing of Al (2s$^2$, 2p$^6$, 3s$^2$, 3p$^1$) and Sc (3s$^2$, 3p$^6$, 4s$^2$, 3d$^1$) electrons was handled using the Virtual Crystal Approximation (VCA) method (see Ref. [\onlinecite{Bellaiche2000}] for information about VCA). In other words, we are modelling Al$_{1-x}$Sc$_{x}$N solid solutions by the means of virtual $<C>$N systems whose cation, $C$, is a compositional average between a $(1-x)$ ratio of Al and a $x$ ratio of Sc. As we are going to see, such averaging constitutes  a rather good approximation for structural and polar properties of Al$_{1-x}$Sc$_{x}$N alloys. It also makes DFT calculations and analyses much simpler than treating large supercells inside which real Al and Sc ions are distributed.

\section{Notations to be adopted, structures to be studied and analytical expression of their energies}

Let us now consider phases that adopt an hexagonal structure, with (i) lattice vectors given by ${\bf a}=a(\frac{1}{2} {\bf x} - \frac{\sqrt 3}{2} {\bf y}$), ${\bf b}=a(\frac{1}{2} {\bf x} + \frac{\sqrt 3}{2} {\bf y}$) and ${\bf c} = c {\bf z}$; and with $<C>$ ions located at ${\bf r}_1$ and ${\bf r}_2$, and N ions placed at ${\bf r}_3$ and ${\bf r}_4$ in the primitive cell with:
${\bf r}_1={\bf 0}$, ${\bf r}_2=\frac{2}{3} {\bf a} + \frac{1}{3} {\bf b} + \frac{1}{2} {\bf c}$, ${\bf r}_3=u {\bf c}$, ${\bf r}_4=\frac{2}{3} {\bf a} + \frac{1}{3} {\bf b} + (\frac{1}{2} +u) {\bf c}$.
Here, {\bf x}, {\bf y}  and {\bf z} are the unit vectors along the Cartesian, x-, y- and z-axes, respectively. $a$ and $c$ are two different lattice parameters with their $c/a$ ratio being called the axial ratio. Finally, $u$ is an internal parameter, that is dimensionless.

Such general construction can be used to describe different phases but not the rocksalt structure (because it is cubic in symmetry) that is known to appear in (Al,Sc)N systems for larger Sc compositions and that will not be investigated in the present work. Examples include the hexagonal structure for which the $c/a$ axial ratio is typically close to 1.2 while the $u$ internal parameter is equal to 0.5. This phase is denoted here by H5 and is shown in Fig. 1a. It is 5-time coordinated since it possesses 5 Al/Sc-N nearest-neighbor bonds of equal distance. It is paraelectric,  with a 6/mmm point group (see, e.g. Ref. \cite{Farrer}). Another case is the wurtzite phase, to be abbreviated as WZ and that is characterized in its ideal form by an axial ratio $c/a$=$\sqrt{\frac{8}{3}}$ $\simeq$ 1.633 and an internal parameter $u=\frac{3}{8}=0.375$. The WZ state is depicted in Fig. 1c, is 4-time coordinated and polar, with a  6mm point group  (see, e.g., Ref. \cite{Farrer} and references therein). One can also imagine other hexagonal structures  for which three bonds are shorter than the purely out-of-plane bond, which we denote as H3+1 and which can happen if, e.g.,  $c/a$ $\simeq$ 1.47 and $u$ $\simeq$ 0.42  -- as shown in Fig. 1d. Structures having the reverse hierarchy, that is the out-of-plane bond being now shorter than the other three nearest bonds, also belong to that class of hexagonal states. We denote them as H1+3 and  one example is displayed in Fig. 1e for the case of $c/a$ $\simeq$ 1.74 and $u$ $\simeq$ 0.34. Note that H3+1 and  H1+3 will be mentioned in Section IV.D. Furthermore, this general hexagonal structure also applies to a phase that we will take as our reference in energy and that is such as (1) its $c/a$ is half-way between that of the ideal H5  structure and the one of the ideal wurtzite phase, that is $c/a$=(1.633+1.2)/2=1.416; (2) $u=0.5$, therefore implying that the reference phase is paraelectric; and (3) its lattice parameters a=b and c are those minimizing the internal energy when imposing $c/a$=1.416 and $u$=0.5. Such a reference structure is displayed in Fig. 1b. Note that H3+1 and H1+3 exhibit the 6mm point group, as the WZ state, while the reference structure possesses the 6/mmm point group like the H5 state.

 \begin{figure}
    	\centering
    	\includegraphics[scale=1.0]{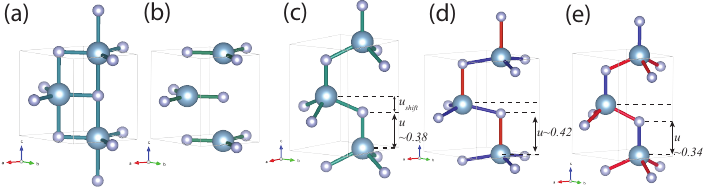}
    	\caption{Different hexagonal phases of Al$_{0.8}$Sc$_{0.2}$N systems, according to our VCA-DFT computations: (a) H5, (b) reference, (c) wurtzite, (d) H3+1 and (e) H1+3  structures.  Shorter bonds are indicated in blue while longer bonds are shown in red in Panels (d) and (e).\label{fig-1}}
\end{figure}

The $a=b$ and $c$ of this reference structure, to be called $a_{ref}$ and $c_{ref}$, predicted by our VCA-DFT computations are displayed in Table 1  for the studied (Al,Sc)N  systems. Note that $a_{ref}$ and $c_{ref}$ depend on the Sc composition, and that we consider here only concentrations of Sc equal to or below  40\% in (Al,Sc)N since these are the compositions that have been shown to be most relevant to experiments in the sense that beyond such compositions the solid solutions prefer to adopt the cubic, non-polar rocksalt structures. \cite{Satoh,Akiyama}.

\begin{table*} 
	\centering
		\caption{Lattice parameters (in \AA) of the Reference structure for  some Al$_{1-x}$Sc$_{x}$N VCA systems, as predicted by our VCA-DFT calculations}
		\begin{tabular}{cccccclllllll}
	\hline\hline
	    Lattice Parameter   & AlN & Al$_{0.9}$Sc$_{0.1}$N  &  Al$_{0.8}$Sc$_{0.2}$N & Al$_{0.7}$Sc$_{0.3}$N & Al$_{0.6}$Sc$_{0.4}$N  \\
	    \hline
		 a$_{ref}$  & 3.227 &  3.2799 &    3.3400 & 3.3893 & 3.4067  & & \\
		\hline
               c$_{ref}$  & 4.5605 & 4.6443 &   4.7294 & 4.7993  & 4.8239  & & \\
               \hline\hline
	\end{tabular}
	\label{tab1}
 \end{table*}

Let us also introduce another internal parameter simply defined as $u_{shift}=0.5-u$. Any $u_{shift}$ different from zero characterizes a polar system, since $u_{shift}$ should be directly proportional to the electrical polarization.

Let us now indicate the following internal energy that is valid for any phase adopting the aforementioned general hexagonal structure, according to  symmetry arguments and group theory:

\begin{equation}
E_{tot} = E_{polar}  + E_{elastic}  + E_{int} 
\end{equation}

with 

\begin{equation}
E_{polar} =  E_{ref}  + \kappa u^2_{shift} +  \alpha u^4_{shift},  
\end{equation}

 \begin{align}
E_{elastic}  & =   s_{3,3,1} \eta_{H,3,3}  + s_{1,1,1} (\eta_{H,1,1} + \eta_{H,2,2}) \\ \nonumber
& + s_{3,3,2} \eta^2_{H,3,3}  + s_{1,1,2} (\eta^2_{H,1,1} + \eta^2_{H,2,2})  + s_{1,3,2} (\eta^2_{H,1,3} + \eta^2_{H,2,3}) +   s_{1,2,2} \eta^2_{H,1,2} \\ \nonumber
& + s_{1,1,3,3} (\eta_{H,1,1} + \eta_{H,2,2})\eta_{H,3,3}   + s_{1,1,2,2} \eta_{H,1,1} \eta_{H,2,2}  + s_{1,3,2,3} \eta_{H,1,3} \eta_{H,2,3} \\ \nonumber
& + s_{3,3,3} \eta^3_{H,3,3} + s_{1,1,3} (\eta^3_{H,1,1} + \eta^3_{H,2,2})  + s_{1,1,2,2,2,1}  (\eta^2_{H,1,1}\eta_{H,2,2} + \eta_{H,1,1} \eta^2_{H,2,2})    \\ \nonumber 
& + s_{3,3,4} \eta^4_{H,3,3} + s_{1,1,4} (\eta^4_{H,1,1} + \eta^4_{H,2,2})     \\ \nonumber
\end{align}

and 
\begin{equation}
E_{int} =  B_{3,1,2}  u^2_{shift}  (\eta_{H,1,1} + \eta_{H,2,2}) + B_{3,3,2}  u^2_{shift}  \eta_{H,3,3}   
\end{equation}

Here, $\eta_{H,i,j}$ are the elements of the homogeneous strain tensor, with the 1, 2 and 3 axes lying along  {\bf a}, {\bf b} and {\bf c} of the hexagonal system, respectively.  By definition of a reference structure,  the zero of strain for  $\eta_{H,1,1}$ and  $\eta_{H,2,2}$ corresponds to the lattice parameter $a$=$b$=$a_{ref}$ while  $c$=$c_{ref}$~  is associated with the zero of strain for  $\eta_{H,3,3}$.
Moreover, a null value for  $\eta_{H,1,2}$ is linked to an angle of 120$^0$ between {\bf a} and {\bf b}. On the other hand, $\eta_{H,1,3}$=0 corresponds to an angle of 90$^0$ between the {\bf a} and {\bf c} lattice vectors, and the zero of strain for  $\eta_{H,2,3}$ is also associated with an angle of 90$^0$, but between {\bf b} and {\bf c}.

 $E_{polar}$ comprises energetic terms related to the polarization up to the fourth order in $u_{shift}$, with  $\kappa$ and $\alpha$ being the parameters appearing in Eq. (2).  Furthermore, $E_{elastic}$ gathers the elastic energies up to the second order in shear strains and fourth order in the diagonal elements of the strain tensor, with the $s$ coefficients 
of Eq. (3) quantifying such elastic energies. Note that the need to involve these fourth orders originates from the fact that the in-plane $a$ and out-of-plane $c$ lattice constants considerably vary between H5 and WZ within the same system, as we are going to see later on. Finally, $E_{int}$ characterizes the coupling interactions between  $u_{shift}$ and the homogeneous strains, with the $B$'s parameters of Eq. (4) representing the strength of such couplings. Technically,  all the coefficients appearing in Eqs. (2), (3) and (4) can depend on the Sc composition in (Al,Sc)N compounds. The Landau-Devonshire model of Ref. \cite{Kei-Andriy} used similar degrees of freedom and decomposition of the internal energy into these three main energies to investigate wurtzite  Al$_{1-x}$Sc$_{x}$N, with the coupling between polarization and strains being similar to ours. On the other hand, their polar energy was pushed up to sixth order in polarization, and, more importantly, their elastic energy only contained second-order in strains. In fact, the hexagonal symmetry of ferroelectric nitrides also allows linear terms in strains in Eq. (3) when selecting some states for reference (like the one of Fig. 1b), which contrasts with the case of ferroelectric perovskite oxides for which the high-temperature paraelectric and cubic phase is typically taken as reference \cite{Landau}. As demonstrated below, these linear terms are important to describe hexagonal ferroelectric nitrides.

\section{Results}

\subsection{Equilibrium wurtzite and H5 structures and energy path from DFT}

  Table 2 displays the lattice vectors and internal parameters $u$ of the WZ and H5 structures for the studied compositions in (Al,Sc)N systems, resulting from the minimization of their total energy within our VCA-DFT computations. The  $c/a$ and $u_{shift}$ of the wurtzite structure significantly varies with Sc compositions, namely from 1.603 to 1.5367 and from 0.1185 to  0.1064, respectively, when the Sc concentration increases from 0\% to 40\%. At the same time,  the axial ratio decreases from $\simeq$ 1.264 to 1.201 and the internal parameter is always null for the H5 structure, when the Sc composition varies between 0\% and 40\%.
  
 In order to check if the use of the VCA scheme is appropriate for describing hexagonal and disordered  (Al,Sc)N  solid solutions, Figs 2a and 2b compare its predicted $c/a$ and $u_{shift}$ of the WZ phase with those arising from the use of  Special Quasi-random Structures (SQS)~\cite{Zunger1990}  having real Al, Sc and N atoms within DFT. Technically, 
 these SQS  all possess a $2\times 2 \times 3$ wurtzite supercell, and are generated thanks to the \textsc{Alloy Theoretic Automated Toolkit}~\cite{vandeWalle2013}. DFT calculations using the Vienna Ab Initio Software package~\cite{Kresse1994,Kresse1996} with Projector Augmented Waves (PAW) pseudopotentials~\cite{Blochl1994,Kresse1999} and the PBE exchange-correlation~\cite{Perdew1996} were conducted on these SQS. We set a plane wave cut-off of 620~eV and a $\Gamma$-centered $10\times10\times4$ sampling of the first Brillouin zone. The electronic self-consistent cycle was considered converged when differences in energy were lower than $10^{-8}$~eV, and the structure was relaxed until forces were smaller than $10^{-3}$  eV/\AA.

Figures 2 demonstrate an excellent agreement for the axial ratio and  $u_{shift}$ internal parameter between the VCA approach and the use of SQS structures, both within DFT, for any studied composition. Moreover, Fig. 2a also reports the experiments of Refs. \cite{Satoh} and \cite{Akiyama}, which were able to measure axial ratio for Sc concentrations up to 38\% and 45\%, respectively. Figure 2 further reveals that our presently used VCA approach can reproduce well the observed $c/a$ of Al$_{1-x}$Sc$_x$N up to $\simeq$ 20\% of Sc concentration. Moreover, one can also estimate the polarization predicted by the present VCA approach by being equal to $\frac{4Z^* u_{shift}}{\sqrt3 a^2_{WZ}}$, where $Z^*$ is the magnitude of the Born effective charge of N ions in the wurtzite phase and $a_{WZ}$ is the lattice parameter of that phase. Using the VCA data of Table 2 for Al$_{0.8}$Sc$_{0.2}$N, along with  $Z^*$=3.2e (as given by DFT calculations in AlScN systems, and where $e$ is the elementary charge) yields a polarization equal to  
128 $\mu C$/$cm^2$, which remarkably agrees with the measurement of Refs. \cite{Zhu,Drury}  providing values of 120-140 $\mu C$/$cm^2$  for Al$_{1-x}$Sc$_x$N compounds with $x$ close to 20\% and also with the value of 124 $\mu C$/$cm^2$ obtained by using the Berry-phase approach \cite{Domenic2,Resta} within DFT and SQS for a Sc composition of 20.8\%.

 It is therefore clear that one can trust the VCA scheme within DFT and the use of $\frac{4Z^* u_{shift}}{\sqrt3 a^2_{WZ}}$ for modelling well structural properties and polarization of disordered Al$_{1-x}$Sc$_x$N systems with $x$ ranging from 0 to 0.2, even if Al and Sc do not belong to the same column of the Periodic Table. This good modeling within VCA may arise from the fact that Al$^{+3}$ and Sc$^{+3}$ have similar ionic radius. For instance, when six-times coordinated,  Al$^{+3}$ and Sc$^{+3}$ have an ionic radius equal to 0.535 and 0.745  \AA, respectively \cite{ionicradius}. For comparison, the ionic radius of La$^{+3}$ is 1.032 \AA, which will make doubtful the use of VCA for (Al,La)N systems).

What is also interesting to realize is that the experimental axial ratio for higher Sc compositions are much lower than the ones predicted by DFT (using VCA or SQS structures). For instance, for a Sc concentration of 0.40, computations provides a $c/a$ of about 1.53-1.54 while the measurements of Ref. \cite{Akiyama} give a value of $\simeq$ 1.47. We thus would like to raise the possibility that,  for $x$ equal or larger than 30\% and up to $\simeq$ 45\%, the wurtzite phase may not be the ground state but rather a metastable state that can coexist with either the H5 structure or even more likely the rocksalt phase that have both much smaller $c/a$ -- namely, of about 1.2 and 1, respectively. Such a hypothesis may be in line with the fact that both Refs. \cite{Satoh} and \cite{Akiyama} found that the observed in-plane lattice parameter $a$ increases much faster with Sc composition for $x$ larger than 30\% and that the out-of-plane lattice parameter $c$ measured in Ref. \cite{Akiyama} decreases with Sc concentration above 30\% while it was increasing when enhancing $x$ from 0 to 0.2 (note that 
the unit cell volume extracted from Ref. \cite{Akiyama} as a function of Sc also deviates from the Vegard's law at 30\% and that a non-monotonic behavior of $c$ with compositions was also observed in Ref. \cite{Andryinew}). Experiments are called for to confirm such a possibility

 \begin{figure}
    	\centering
    	\includegraphics[scale=0.6]{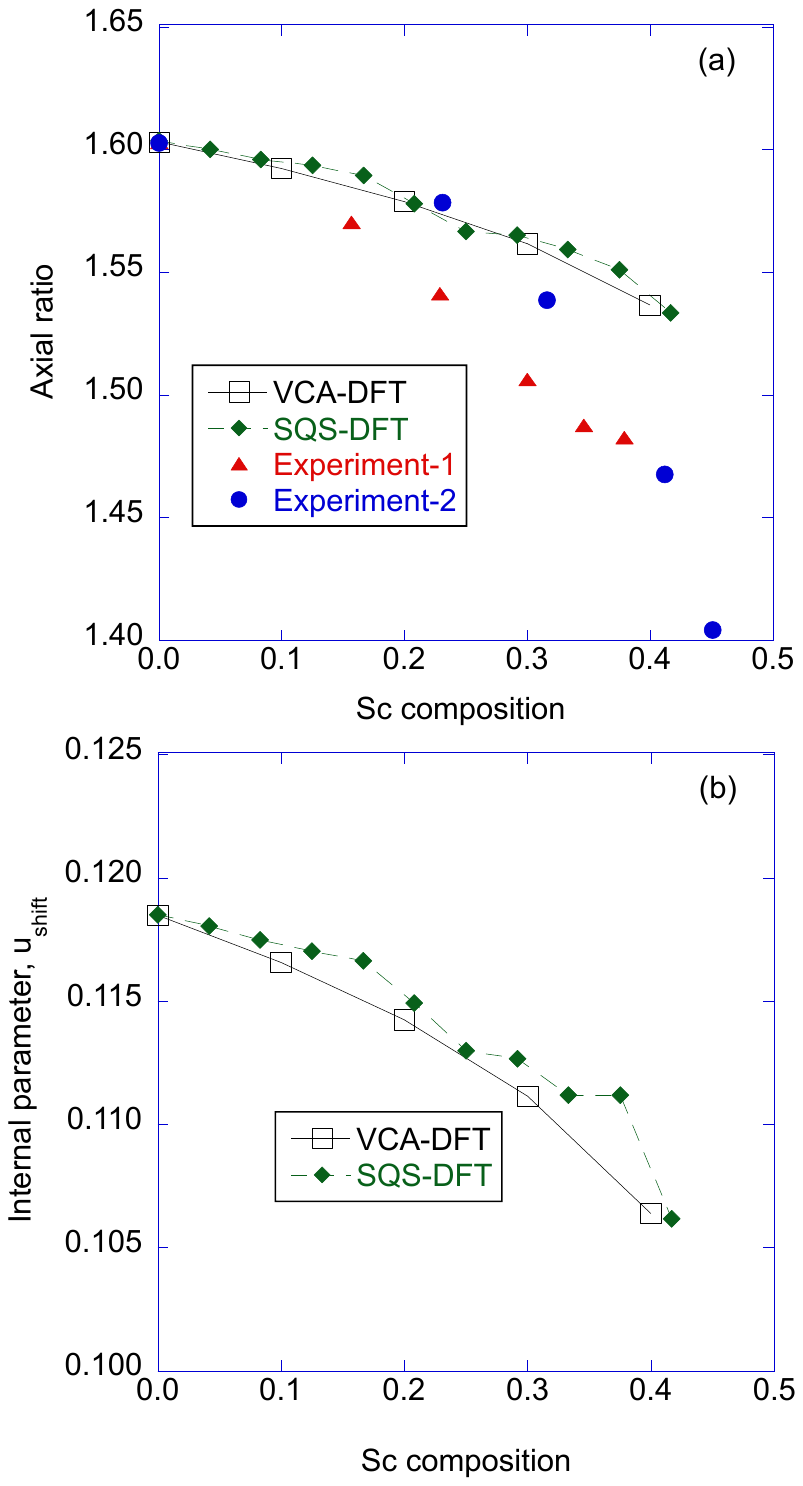}
    	\caption{$c/a$ (panel a)  and $u_{shift}$ (panel b) {\it vs.} composition in some (Al,Sc)N systems. VCA-DFT  correspond to the  VCA approach with DFT, while  SQS are employed within DFT for the data labelled SQS-DFT. Experiment-1 and Experiment-2 are measurement of Refs. \cite{Satoh} and \cite{Akiyama}, respectively.  \label{fig0}}
\end{figure}

\begin{table*} 
	\centering
		\caption{Lattice parameters and internal parameters of the wurtzite (WZ) and hexagonal H5  phases  for  the studied Al$_{1-x}$Sc$_{x}$N systems, as predicted by our VCA-DFT systems} 
	\begin{tabular}{cccccclllllll}
	\hline\hline
	    Coefficient   & AlN & Al$_{0.9}$Sc$_{0.1}$N  &  Al$_{0.8}$Sc$_{0.2}$N & Al$_{0.7}$Sc$_{0.3}$N & Al$_{0.6}$Sc$_{0.4}$N  \\
	    \hline
	    \hline
		$a=b$ of WZ  (\AA) & 3.1268  & 3.1930  & 3.2551 &  3.3140   &  3.3726   \\
		\hline
		 $c$ of WZ (\AA) & 5.0124  & 5.0837 & 5.1388 & 5.1753 & 5.1827   \\  
		\hline
		 $c/a$ of WZ   & 1.6030  & 1.5921 & 1.5787 &  1.5616 & 1.5367    \\  
		\hline
		internal parameter, $u$, of WZ    &  0.38153 & 0.38342 & 0.38576 &  0.38885 &  0.39361     \\ 
		\hline
		internal parameter, $u_{shift}$, of WZ   &0.11847  & 0.11658 &  0.11424 & 0.11115   & 0.10639   \\ 
		\hline
		\hline
		$a=b$ of H5  (\AA) & 3.3107  & 3.3899  & 3.4587 &  3.5193  & 3.5715   \\
		\hline
		 $c$ of H5  (\AA) & 4.1830 & 4.1959  & 4.2213 & 4.2522 & 4.2885    \\  
		\hline
		 $c/a$ of H5  &1.2635  & 1.2378  & 1.2205 & 1.2083 & 1.2008   \\  
		\hline
		internal parameter, $u$, of H5  &  0.500 & 0.500 &  0.500 &  0.500 &  0.500     \\ 
		\hline
		internal parameter, $u_{shift}$, of H5  & 0.000  & 0.000 &  0.000  & 0.000   & 0.000    \\ 
	        \hline\hline
	\end{tabular}
	\label{tab2}
\end{table*}

 Let us now pay attention to an energy path connecting the equilibrium H5 structure to the equilibrium WZ phase for our studied compositions as predicted by our VCA-DFT computations. For that, we consider structures for which the $u$ internal parameter as well as the $a$ and $c$ lattice parameters are linearly interpolated and extrapolated between these two phases, and compute their energy. Figures 3(a) and 3(b)  show the resulting energy {\it versus} $c/a$ and as a function of $u_{shift}$, respectively, for our various (Al,Sc)N  alloys, choosing the zero of energy as the energy of the (compositionally-dependent) reference structure (because of the aforementioned interpolation and extrapolation, each data point in Figs. 3a and 3b is associated with different   ($u_{shift}$,$c/a$) combination. In other words, $u_{shift}$ evolves when $c/a$ changes and {\it vice-versa}).

\begin{figure}
    	\centering
    	\includegraphics[scale=0.7]{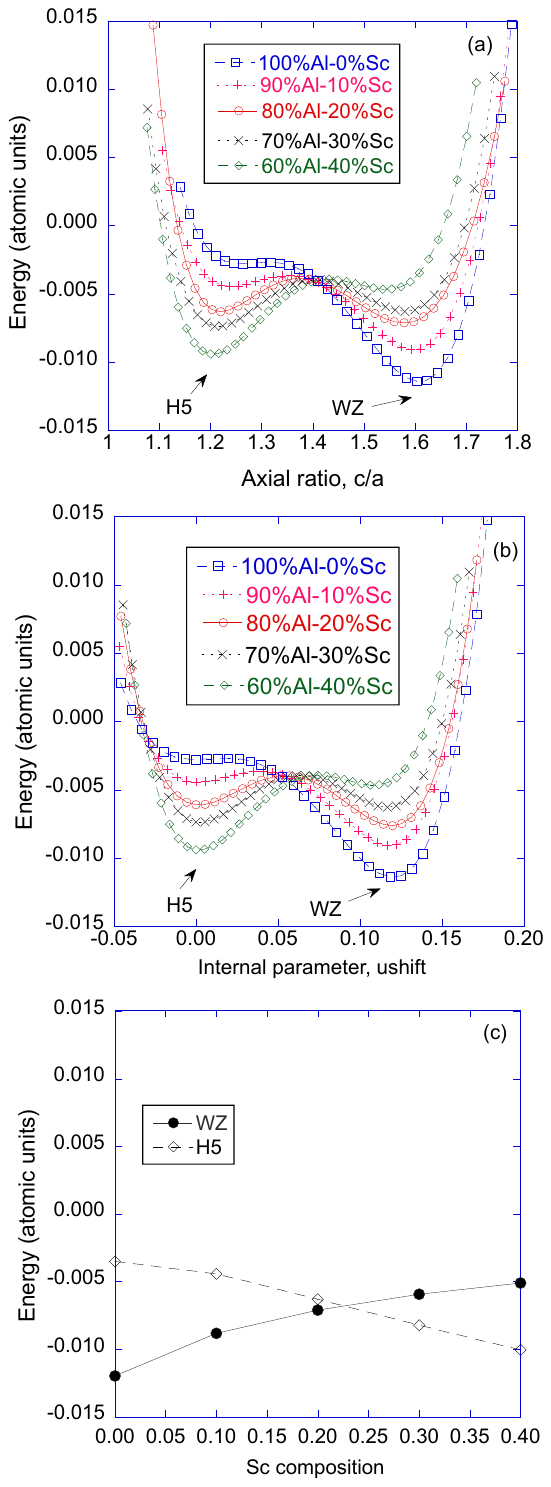}
    	\caption{Some energies: Energy versus $c/a$ (panel a), versus $u_{shift}$ (panel b), and of WZ and H5 structures versus composition (panel c), in some (Al,Sc)N systems and as predicted by our VCA-DFT approach. The zero of energy is chosen to be the one of the reference structure.    \label{fig1}}
\end{figure}

The energy minimum of the wurtzite structure becomes shallower when increasing the Sc composition, even basically almost disappearing for the Sc concentration of 40\% (note that we numerically checked that it does disappear for higher Sc compositions, such as 50\%). In contrast,  the H5 structure possesses an energy that becomes deeper, even ``overtaking'' that of the wurtzite structure between 20\% and 30\% of Sc, as emphasized in Fig. 3c (note, however, that we numerically found that using SQS structures and different functionals within DFT can lead to a slightly higher Sc critical composition at which the H5 structure has a lower energy than  WZ). Such latter fact emphasizes again a possibility that we raised above, namely that the WZ phase is not the ground state anymore in disordered Al$_{1-x}$Sc$_x$N solid solutions with Sc compositions larger than $\simeq$ 30\%; the energies shown in Fig. 3c, and more precisely their difference,  quantify the relative stability of the WZ {\it versus} the H5 phase for each selected composition.

 \subsection{Procedures to get the parameters of the Landau-type model described by Equations (1)-(4)}

We now want to take advantage of Eqs. (1)-(4) to better understand (Al,Sc)N compounds and the aforementioned DFT results.
For that, we first need to extract all the parameters of Eqs (2)-(4), which we do by conducting different and specific additional VCA computations within DFT.
The procedure we adopted has three main steps. 

First of all, we perform VCA-DFT simulations to determine these parameters one-by-one. For instance, starting from the reference structure and only varying $u_{shift}$ by hand and then obtaining the resulting energy curve as a function of $u_{shift}$ allows to fit such curve by Eq. (2) and therefore to extract $E_{ref}$, $\kappa$ and $\alpha$.  Similarly, obtaining the $s_{1,2,2}$  parameter of Eq (3) simply requires to start with the reference structure, but now simply making $\eta_{H,1,2}$ finite and ranging between different values. The generated energies should  obey  $E_{0}$ +  $d \eta_{H,1,2}^2$ , with $E_{0}$ being $E_{ref}$ too and the fitting $d$ coefficient corresponding to  $s_{1,2,2}$. For the determination of $s_{1,3,2}$, this can be done by starting from the reference structure again but now making  $\eta_{H,1,3}$ deviating from zero and varying within a range. The resulting energies should satisfy $E_{0}$ +  $e \eta_{H,1,3}^2$, with $E_{0}$ being $E_{ref}$ as well and the free $e$ parameter being  $s_{1,3,2}$. Then using again the reference structure as starting point and forcing $\eta_{H,1,3}$ to be equal to $\eta_{H,2,3}$ and be both finite within a window of values yields energies that are given by  $E_{0}$ +  $f \eta_{H,1,3}^2$, with $f$ corresponding to $2s_{1,3,2}$+$s_{1,3,2,3}$. This allows to determine $s_{1,3,2,3}$ as equal to $f$-$2s_{1,3,2}$. Regarding the parameters entering Equation (4), both strains and  $u_{shift}$ have to move away from their vanishing values of the reference structure in DFT calculations. The resulting energies can then be fitted by an energy appearing in Eq.(4) to get the  $B$ coefficients.

In a second main step, we then adjust some of these DFT-determined parameters by also incorporating the first-principle energy paths displayed in Figs 3  into the fit of such coefficients. Finally, the last step consists in slightly varying some of the parameters of Eqs. (2)-(4) in order that all these coefficients basically continuously evolve when changing the Sc composition in (Al,Sc)N systems. Such continuity also implies that one can extract parameters for compositions not studied here by simply interpolating (or even extrapolating)  the coefficients obtained for other concentrations.

Tables 3 report the resulting extracted parameters of Equations (2)-(4) for our investigated (Al,Sc)N  compounds. We will emphasize important results related to some of these coefficients later on.

\begin{table*} 
	\centering
		\caption{Coefficients (in Hartree/f.u) of Eqs. (2)-(4) for  the studied Al$_{1-x}$Sc$_{x}$N VCA systems} 
	\begin{tabular}{cccccclllllll}
	\hline\hline
	    Coefficient   & AlN & Al$_{0.9}$Sc$_{0.1}$N  &  Al$_{0.8}$Sc$_{0.2}$N & Al$_{0.7}$Sc$_{0.3}$N & Al$_{0.6}$Sc$_{0.4}$N  \\
	    \hline
		$E_{ref}$ & -79.247  & -76.056  & -73.049 &  -70.233   &  -67.613   \\
		\hline
		 $\kappa$ & -1.774  & -1.614  & -1.495 &  -1.430 & -1.364    \\  
		\hline
		 $\alpha$   &  126.016 & 124.052 & 121.155 &  121.818 &  126.547     \\ 
		\hline
		$s_{3,3,1}$  & 0.037  & 0.051 &  0.063 & 0.071   & 0.075    \\ 
		\hline
		$s_{1,1,1}$ & -0.030  & -0.036 &  -0.037 & -0.035   & -0.068   \\ 
		\hline
		$s_{3,3,2}$ &  0.300 & 0.252 & 0.124 & -0.006   & -0.015    \\ 
 		\hline
		$s_{1,1,2}$  &  1.231 & 1.229 & 1.232  & 1.2396 & 1.252   \\ 
		\hline
		$s_{1,3,2}$ &  1.503 & 1.498 &  1.491 &  1.503 & 1.544   \\ 
		\hline
		$s_{1,2,2}$  & 1.955  & 1.937 & 1.863 & 1.893  & 1.966  \\ 
		\hline	
		$s_{1,1,3,3}$  &  0.582 & 0.522 & 0.496 &  0.448 & 0.426  \\ 
		\hline
		$s_{1,1,2,2}$  & 0.391 & 0.440 & 0.627 & 0.813  & 0.895    \\ 
                \hline
		 $s_{1,3,2,3}$  &  0.007 & 0.006 & 0.0041 & 0.002  & -0.000   \\ 
		\hline
		$s_{3,3,3}$  &  -0.161 & 0.073 & -0.238 &  -0.312 & -0.559   \\  
		\hline	
		$s_{1,1,3}$  & -2.943  & -3.061 & -3.302 & -3.702  & -4.302   \\
		\hline 
		$s_{1,1,2,2,2,1}$  &  -0.755 & -0.820  &  -1.471 &  -2.106 &  -2.117  \\ 
		\hline	
		$s_{3,3,4}$  & 9.091  & 11.693 & 13.867 & 16.003  & 18.656   \\ 
		\hline
		$s_{1,1,4}$  & 4.061 & 4.881 & 8.278 & 10.493   &13.554  \\		
		\hline
		$B_{3,1,2}$  &  3.826 & 4.222 &  4.785 &  4.920  & 5.058   \\
		\hline
		$B_{3,3,2}$ &  -11.883 & -12.879 & -12.711 & -12.550 &  -13.310  \\
	        \hline\hline
	\end{tabular}
	\label{tab3}
\end{table*}

 \subsection{Checking the accuracy of the model of Eqs. (1)-(4)}

Meanwhile, let us select the composition of 80\% of Al and 20\% of Sc because Fig. 3  tells us that this is a drastic case in the sense that the minima of the H5 and the WZ structures are not only well pronounced but also have very similar energy. We then plug the inputs of the energy path computed by DFT and shown in Figs 3, that are the values of the strain components and of $u_{shift}$ for any data point in this energy path, into Eqs. (2), (3), and (4) to obtain the corresponding energies predicted by the model described in section III. Figures 4a and 4b reveal that the agreement between such a model and the DFT data of Fig. 3 for Al$_{0.8}$Sc$_{0.2}$N  is very satisfactory, especially for axial ratio ranging between 1.15 and 1.76 and for $u_{shift}$ varying between $\simeq$ -0.03 and +0.17. Such an agreement therefore, attests to the validity of Eqs. (1)-(4) and of the aforementioned procedure to determine its coefficients, at least for the Sc composition of 20\%. Such validity was not obvious at all from the start, when realizing that along the path going from H5 to WZ, (i) a bond is broken since the system goes from a 5-time coordinated structure to a 4-time coordinated phase; and (ii) both the axial ratio and internal parameter $u$ dramatically vary. It is in fact remarkable that the same energy functional of Eq. (1)-(4) can describe structures that are so different.

\begin{figure}
    	\centering
    	\includegraphics[scale=0.7]{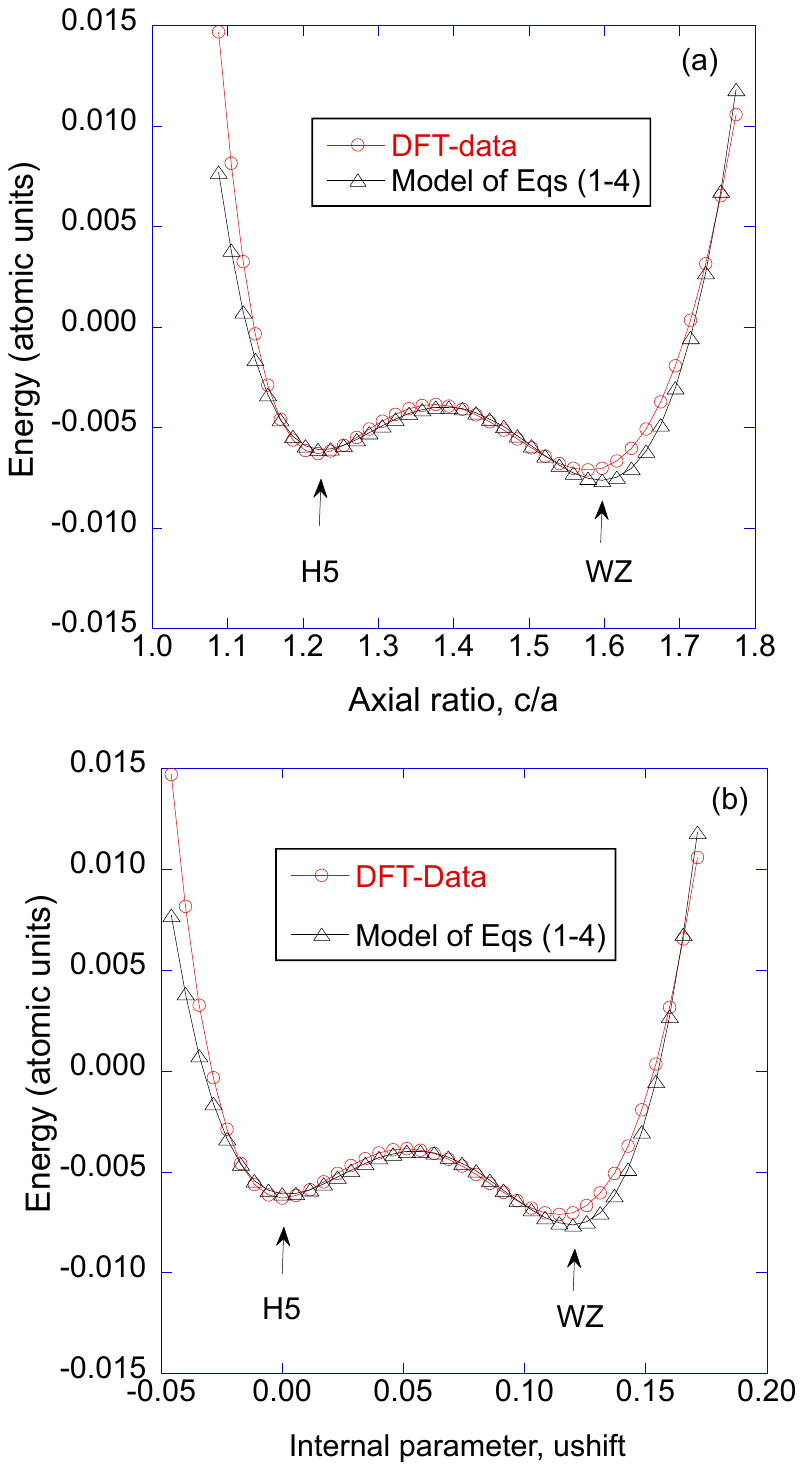}
    	\caption{Energy versus $c/a$ (panel a) and versus $u_{shift}$ (panel b)  in Al$_{0.8}$Sc$_{0.2}$N, as computed from DFT-VCA and as obtained by the Model of Eqs (1)-(4)    \label{fig2}}
\end{figure}

To check if this model is also satisfactory for other concentrations,  Figs. 5a and 5b also depict the energy path predicted by this model and its coefficients indicated in Table 3, as a function of the axial ratio and $u_{shift}$ for {\it any} Sc composition studied here. Comparing Figs 5 and Fig.3 indicates that all the qualitative but also quantitative features of these solid solutions are well reproduced by this model, including the compositional behavior of the axial ratio, internal parameter and energy of both the H5 and WZ phases, and any structure in-between. This model and its compositionally-dependent coefficients can therefore describe well  properties of hexagonal (Al,Sc)N ferroelectric nitrides having different compositions.

\begin{figure}
    	\centering
    	\includegraphics[scale=0.7]{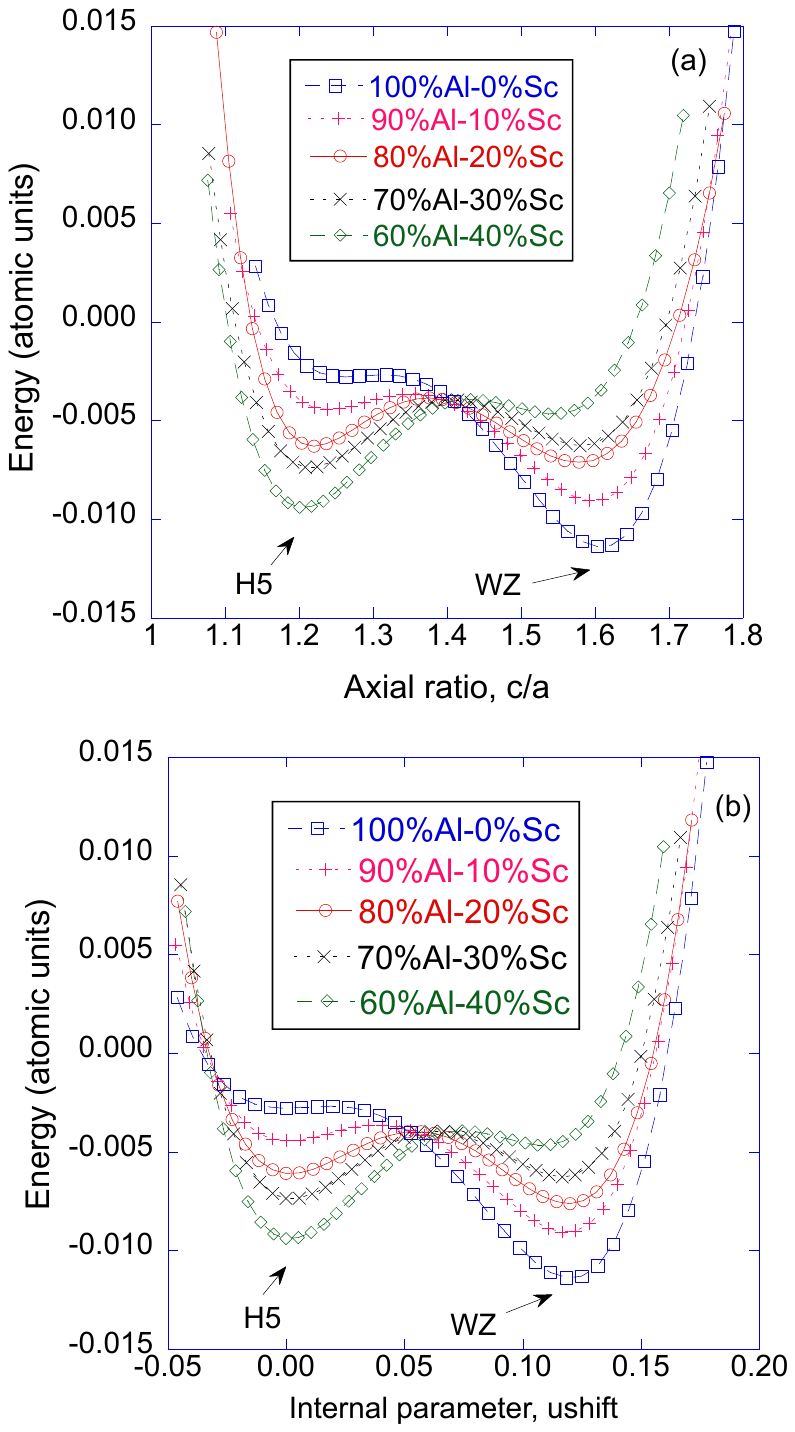}
    	\caption{Energy versus $c/a$ (panel a)  and versus $u_{shift}$ (panel b)  in some (Al,Sc)N systems, as obtained by the Model of Eqs (1)-(4)    \label{fig3}}
\end{figure}

\subsection{Using the Model of Eqs. (1)-(4) to better understand ferroelectric nitrides}

Consequently, let us take advantage of such a model to provide a microscopic insight into (Al,Sc)N  systems.  For that, we first come back to the challenging case of 80\% of Al and 20\% compositions. Figure 6 displays the energy as a function of $u_{shift}$ as predicted by the aforementioned model  {\it  for six different cases}: (i) all the parameters of this model are included (black solid lines); (ii) only the energy solely related to $u_{shift}$, i.e., $E_{polar}$  of Eq. (2), is switched on (blue empty circles); (iii) only the strains are taken into account, that is only $E_{elastic}$ of Eq. (3) is finite (green empty circles); (iv) both $E_{polar}$  of Eq. (2) and $E_{elastic}$ of Eq. (3) are activated (green empty rhombus), implying that the degrees of freedom related to internal parameters and lattice parameters are incorporated but {\it not their couplings described by $ E_{int} $ of Eq. (4)}; (v) the strains are involved as well as their couplings with $u_{shift}$ but not the energetics solely associated with the internal parameter (green crosses). In other words, $E_{elastic}$ of Eq. (3) and $ E_{int} $ of (4) are turned on while $E_{polar}$ of Eq. (2) is turned off; and (vi) the polar displacements along with their coupling with strains are taken into account, but not the elastic energy {\it per se} (bue empty triangles) -- that is $E_{polar}$ of Eq. (2) and $E_{int}$ of (4) are on while $E_{elastic} $ of Eq. (3) is turned off. Figure 6 is very instructive to start to better understand (Al,Sc)N.

One can for instance, notice that only using $E_{polar} $ of Eq. (2) can {\it not} generate the H5  structure (characterized by  $u_{shift}$ close to zero) and also the wurtzite phase (that possesses a $u_{shift}$  near 0.12), as equilibrium or even metastable states. It rather yields a single minimum for which $u_{shift}$ adopts an intermediate value of $\simeq$ 0.08 (and an axial ratio of about 1.47, not shown here). This is precisely the H3+1 structure shown in Fig. 1d.

On the other hand, one can describe energetics of H5 by only taking into account the strain degrees of freedom and their energetics described by $E_{elastic}$ of  Eq. (3), revealing that the H5 structure is ferroelastic in nature. In fact, a relative good agreement is obtained with the full model also {\it near} (and not only at) such H5 phase, that is when $u_{shift}$ varies between $\simeq$ -0.03 and $\simeq$ +0.03. It is important to realize that both $u_{shift}$ and the strains are varying for all the data point shown in Fig. 6, which explains why $E_{elastic}$ is not a horizontal line in Fig. 6.

Adding then the energy related to  $u_{shift}$ to the energy associated with strains now allows a good description even further away from H5, in the sense that using $E_{elastic}$ and $E_{polar}$ of Eqs.(3) and (2) together results in energy being very close to that of the full model for $u_{shift}$ up to $\simeq$ +0.07. However and strikingly, employing $E_{elastic}$ of Eq. (3) alone or in addition to $E_{polar}$ of Eq. (2) can {\it not} induce the wurtzite minimum.

In other words, it is imperative that $u_{shift}$ and the strains are coupled to each other, or equivalently, $E_{int}$  of Eq. (4) needs to be activated in addition to $E_{polar}$ and $ E_{elastic}$ of Eqs. (2) and (3), in order for the wurtzite state to exist! Such a remarkable result demonstrates the tremendous importance of coupling between optical and acoustic modes in nitride ferroelectrics for the wurtzite phase to exist -- the former being related to $u_{shift}$ while the latter are associated with strains. In that sense, one can think of the WZ phase as being a rare example of a trigger-type  ferroelectric state \cite{triggered1,triggered2} since the coupling between a polar mode and another degree of freedom (strain here) is required to induce this polar 4-time coordinated state.

It is also important to realize that turning on the energetics of strains and their couplings with $u_{shift}$ via $E_{elastic}$ and $E_{int}$, but not activating $E_{polar}$ of Eq. (2), still provides a good description near the H5 structure, that is for $u_{shift}$ ranging between $\simeq$ -0.03 and $\simeq$ +0.03, but yields an energy that is not bounded for the largest $u_{shift}$. One therefore absolutely needs the three energies of Eq. (1) to be able to have the energy minima of both H5 and WZ within a hexagonal symmetry.  

The need of possessing the three energies is further emphasized when looking at the results when $E_{polar}$ and $ E_{int}$ of Eqs. (2) and (4)
are activated but not $E_{elastic}$ of Eq. (3). In that case, not only the H5 minimum disappears but the other and now single minimum is not WZ {\it per se}. It is rather the H1+3 state shown in Fig. 1e, for which both $u_{shift}$ and $c/a$ are larger than in the WZ phase, since they are equal to 0.16 and 1.74 here. 

In order to have a quantitative idea about the importance of these three energies, let us provide their values for the WZ structure, that has a total energy of -0.0076 Hartree per four atoms (with respect to the reference structure) for Al$_{0.8}$Sc$_{0.2}$N according to the model of Equation (1): the polar energy $ E_{polar}$ of Eq. (2) is repulsive with a value of +0.0036 Hartree; the elastic energy $E_{elastic}$ of Eq. (3) is even three time more repulsive with +0.0103 Hartree; and the coupling energy $E_{int}$ of Eq. (4)  between $u_{shift}$ and the strains is the only one that is attractive with a strong value of -0.0215 Hartree. For comparison, the model also predicts a purely $E_{elastic}$ elastic energy for the H5 minimum (as it should be since $u_{shift}$=0 for that minimum, therefore annihilating both $E_{polar}$ and $E_{int}$ of Eqs. (2) and (4)) of that composition with a value of -0.0061 Hartree. It is also interesting to compare the energetic decomposition for the WZ minimum of Al$_{0.8}$Sc$_{0.2}$N with that of a prototype ferroelectric oxide, namely bulk PbTiO$_3$ (PTO). An effective Hamiltonian built in a similar way than that of Ref. \cite{Zhong} provides for the ferroelectric tetragonal ground state of PTO a total internal energy of -0.0047 Hartree per 5 atoms with respect to the cubic paraelectric state. Its decomposition yields an elastic energy of +0.0034 Hartree, that is thus also repulsive but about three times smaller with respect to the WZ phase of the nitride system; a coupling energy between polar degrees of freedom and strains of -0.0066 that is therefore attractive too but also about three times smaller in magnitude than for the WZ state of  Al$_{0.8}$Sc$_{0.2}$N; and a polar energy that is now negative and equal to -0.0015 Hartree, making it therefore attractive in nature, in sharp contrast with the case of our studied ferroelectric nitride in its WZ minimum.

\begin{figure}
    	\centering
    	\includegraphics[scale=0.7]{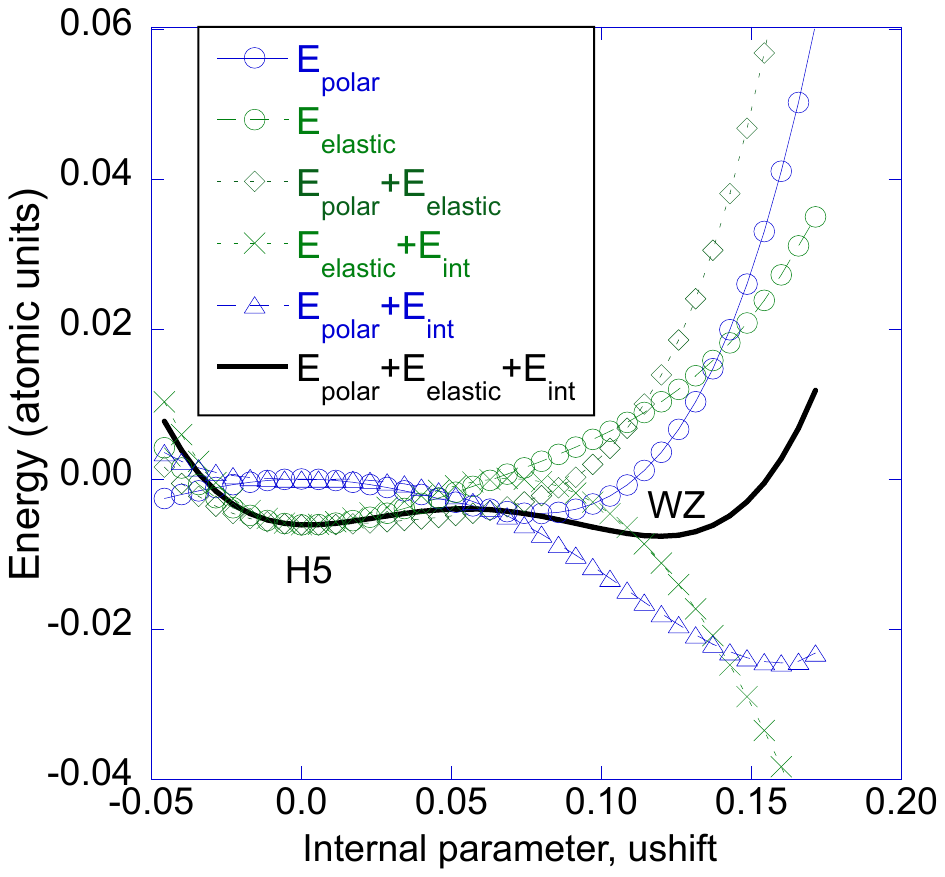}
    	\caption{Energy versus $u_{shift}$ in Al$_{0.8}$Sc$_{0.2}$N, as obtained by the model of Eqs (1)-(4)  when turning off and on some energies  \label{fig4}}
\end{figure}

Let us continue to use such model but now to look for the interactions governing the fact that increasing the Sc composition in (Al,Sc)N makes the relative energy of the wurtzite state moving up while that of the H5 phase becomes deeper, even overcoming that of the wurtzite phase for Sc concentrations of $\simeq$ 30\% and above.
For that, we focus on pure AlN and start with its parameters provided in Table 3. 

We first allow the $\kappa$ coefficient of  Eq. (2) to vary from its pure AlN value of -1.774 to -1.364, by progressively choosing all the $\kappa$ adopted by  Al$_{0.9}$Sc$_{0.1}$N,  Al$_{0.8}$Sc$_{0.2}$N,  Al$_{0.7}$Sc$_{0.3}$N and finally Al$_{0.6}$Sc$_{0.4}$N, when freezing all the other parameters of Eqs (1)-(4) to their AlN values. The resulting energy path is shown as a function of $u_{shift}$ in Fig. 7a, which tells us that increasing $\kappa$ results in the WZ equilibrium state to increase its energy and to adopt a lower $u_{shift}$, precisely as increasing the Sc composition does in (Al,Sc)N  (see  Fig. 3b). However, the energy of the H5 structure is insensitive to that change of  $\kappa$, in contrast to what progressively adding Sc in AlN does. Such insensitivity can be easily understood by the fact that $\kappa$ acts on  $u^2_{shift}$ in Eq. (2) while this latter quantity is null for the H5 structure. 

We then decided to vary another parameter from its value in AlN to that of Al$_{0.6}$Sc$_{0.4}$N, while keeping all the other coefficients frozen to those of pure AlN. This coefficient is a purely elastic one and is therefore only involved in Eq. (3). It is $s_{3,3,1}$, which appears in front of $\eta_{H,3,3}$ there. The results are shown in Fig. 7b, and indicate that such enhancement of  $s_{3,3,1}$ now allows two effects seen in Fig. 3b when increasing the Sc concentration to occur, namely the equilibrium energy of the wurtzite structure is moving up while that of the H5 structure is moving down. However, two undesirable effects {\it not} displayed by Fig. 3b happen as well: (1) the $u_{shift}$ at which the minimal energy of the H5 structure occurs is quite different from zero with its magnitude becoming even bigger when $s_{3,3,1}$ increases from 0.037 (value in AlN) to 0.075 (value in Al$_{0.6}$Sc$_{0.4}$N), and (2) for the largest value $s_{3,3,1}$=0.075, the wurtzite structure is still the minimum in energy. 

One way to overcome these two shortcomings is to vary both $\kappa$ and $s_{3,3,1}$, as demonstrated in Fig. 7c. In other words, varying at the same time these polar and elastic coefficients allows features seen in Fig. 3b when changing composition to occur. In particular, one does not need to tune the $B$ couplings between $u_{shift}$ and the strains to make that happen, despite the fact that Fig. 6 demonstrated that such coupling is essential to have two minima within the hexagonal symmetry. In other words, one message here is that such couplings do need to be taken into account in ferroelectric nitrides but do not necessarily need to be composition-dependent, unlike $\kappa$ and $s_{3,3,1}$ that have to vary with the Sc concentrations  (note that Fig. 7d shows that such dependency is basically linear).

\begin{figure}
    	\centering
    	\includegraphics[scale=0.7]{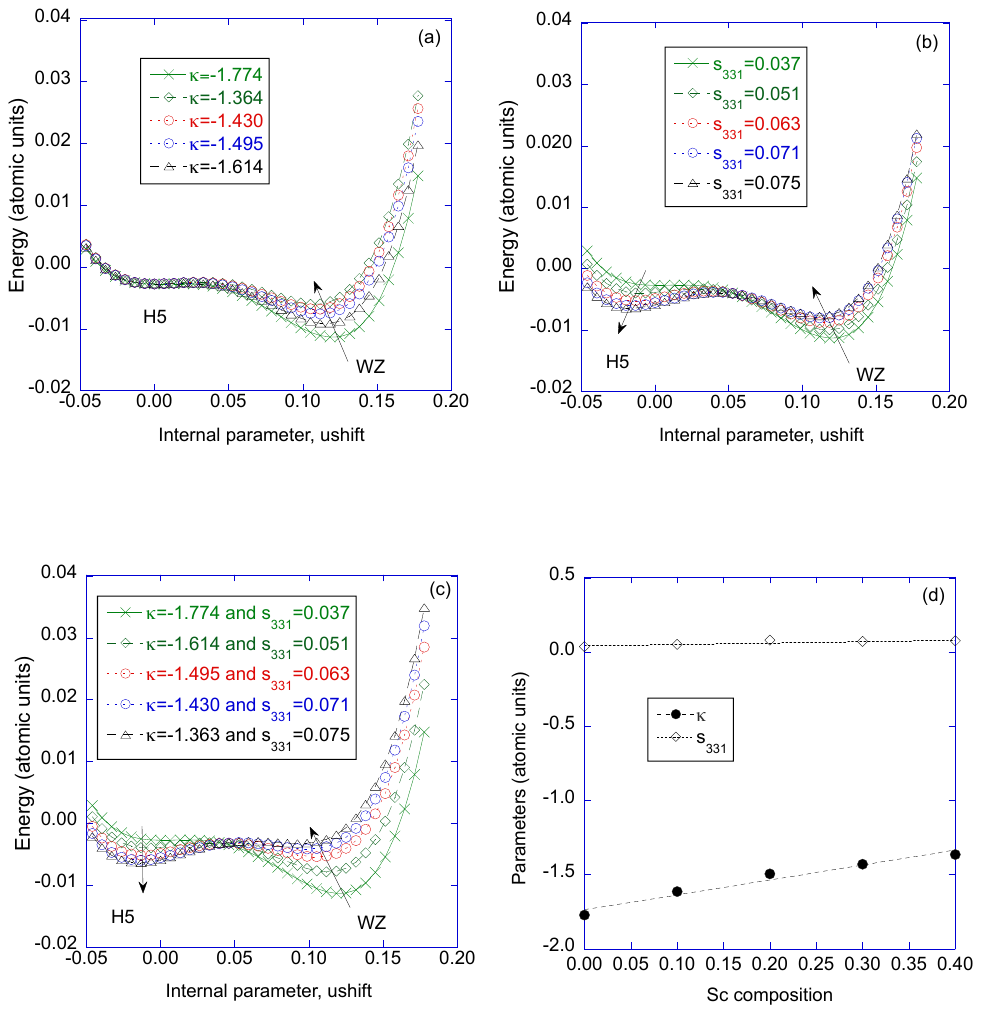}
    	\caption{Energy versus $u_{shift}$ in AlN, as obtained by the model of Eqs (1)-(4)  when varying (a) $\kappa$ of Eq. (2); (b)    $s_{331}$ of Eq. (3)  and (c) both  $\kappa$ of Eq. (2)  and   $s_{331}$ of Eq. (3). Panel d shows the compositional dependency of
	the DFT-extracted $\kappa$ and $s_{331}$  parameters, with the dashed lines representing linear fits. \label{fig5}}
\end{figure}

\section{Perspectives}

One may also wonder if the model of Eq. (1)-(4) and its parameters provided in Table 3 can offer other perspectives, in addition to provide the aforementioned deep insights into nitride ferroelectrics.

\subsection{Looking for hidden states}

It  is indeed the case once realizing that having a parametrized model with relatively simple equations allows one to easily compute the internal energy as a function of both $u_{shift}$ and $c/a$. Here, we assume that the system remains in the hexagonal configuration by setting $\eta_{H,1,2}=\eta_{H,2,3}=\eta_{H,1,3}=0$ and $\eta_{H,1,1}=\eta_{H,2,2}$. Naively, we started by fixing $c/a$, $u_{shift}$ and varying $\eta_{H,1,1}$ (which fixes $\eta_{H,3,3}$) until a minimum in energy is found. However, this has the problem that the function is neither continuous in energy nor in the strain parameters, which becomes non-physical. We then looked for a simpler model that was then found to reproduce the method just described within a difference of $3\times10^{-4}$ Hartree in the path that minimizes the energy (dashed line in Fig. 8). Practically, we started with the reference structure from Tab. 1 and assume that $\eta_{H,1,1}$ varies linearly with $c/a$ with a slope given by $m:=(\eta_{H,1,1,WZ}-\eta_{H,1,1,H5})/((c/a)_{WZ}-(c/a)_{H5})$; In this context, the subscript $WZ$ and $H5$ allude to the parameters that are obtained by using the  WZ and H5 structures laid out in Tab. 2, respectively.
 An example of such results is provided in Fig. 8  for  Al$_{0.7}$Sc$_{0.3}$N, for which the lowest energy in the hexagonal system is now paraelectric H5 rather than polar WZ. Figure 8 employs  a color code to represent the energy. There, the red color emphasizes the states with low energies. One can easily see a plethora of low-energy states not only near the H5 and WZ minima but also in-between with the combinations ($u_{shift}$,$c/a$) varying from $\simeq$ (0.06,1.3) to $\simeq$ (0.10,1.5). These latter states may be considered to be hidden states \cite{15,16,17,18,19,20}, in the sense that they are probably not reachable by thermodynamic conditions but rather only accessible when driving the system out of equilibrium. Such drive and occurrence of these hidden states can occur by, e.g., applying a train of pulses of electric field, starting from the H5 paraelectric state and finally reaching the WZ polar phase by successive activation of hidden states with increasing value of the polarization, as similar to what was recently predicted to happen in complex perovskite oxides having a non-polar initial state and a final strong ferroelectric metastable phase \cite{SergeyPMN,SergeyNNO,SergeyBRFO}. It is important to realize that hidden states can possess better properties than ground states and are also key features to design neuromorphic computing \cite{SergeyPMN,SergeyNNO,SergeyBRFO, SergeyPZTfilms,SergeyReview,Boyn,21}.
 Plots such as Figure 8 can also reveal what paths {\it with low energies} the system under study can adopt when going from the H5 to WZ states, in case that the model of Eqs (1-4) is very accurate (note that the path shown in Fig. 5 is not the minimal energy path because linear interpolations between the $u_{shift}$, as well as $a$ and $c$ lattice parameters, of H5 and WZ minima given in Tab. 2 were assumed there, unlike in the true minimal energy path).  The yellow dashed line reports the minimal energy path found by VCA-DFT computations, which indeed goes through states of low energies predicted by our Model. On the other hand, the blue dashed line, which corresponds to the minimal energy path provided by our model of Eqs (1)-(4) with the parameters given in Table 3, shows that the agreement between our model and DFT-VCA are not perfect for this path, even if both dashed lines demonstrate that these paths starts near H5 with states that prefer to have a lower $c/a$ than the one simply given by a linear interpolation between H5 and WZ. Note that agreement can be further improved by fine-tuning the parameters of Eq.(1) (which is not the aim of this article) and that  it is also worthwhile to know that both dashed lines provide a very similar value for the energy barrier in  Al$_{0.7}$Sc$_{0.3}$N, namely 0.089 eV/f.u. for the blue line and 0.088 eV/f.u. for the yellow line -- which also compares rather well with the energy barrier directly given by DFT-VCA computations and which is equal to 0.099 eV/f.u.


\begin{figure}
    	\centering
    	\includegraphics[scale=0.8]{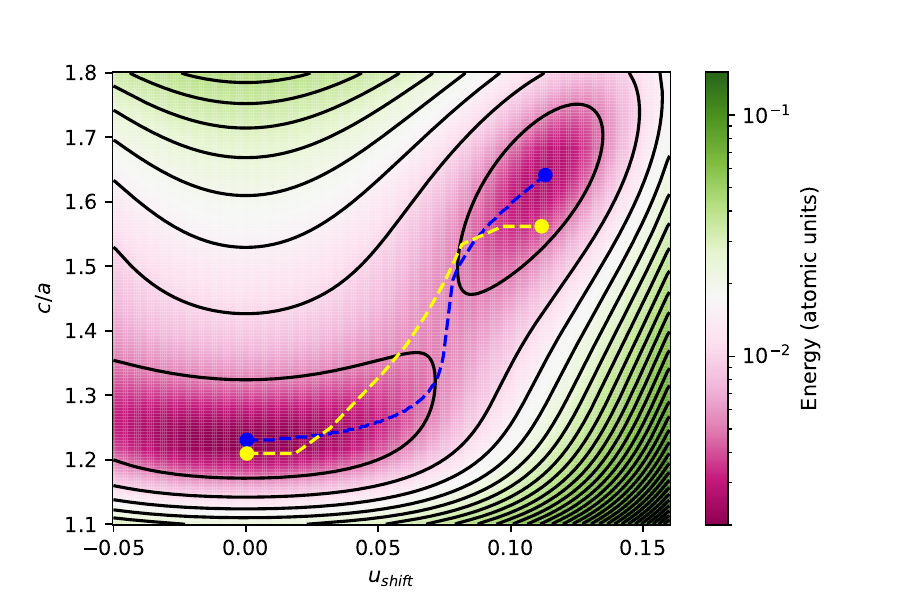}
    	\caption{Energy landscape in the ($u_{shift}$,$c/a$) plane for Al$_{0.7}$Sc$_{0.3}$N, as obtained by the model of Eqs (1)-(4). The zero of energy is chosen here to be able to better appreciate the scale of energy. The blue and yellow dashed lines represent the path of minimal energy to go from the H5 to WZ state according to our model with the parameters of Table 3 and VCA-DFT, respectively. The dots represent the location of the H5 and WZ minima predicted by this model and VCA-DFT. \label{fig6}}
\end{figure}

\subsection{Finite-temperature properties}

Having simulations at finite temperatures is also highly desired in these
nitride ferroelectrics since most devices operate around 300K, and since these
compounds are highly promising for high-temperature electronics as well.
However, in that case, one needs to work with the free energy rather than the
internal energy, as we have done so far when conducting DFT calculations at 0K
or when using Eqs (1)-(4) of the proposed model.

One straightforward way to go from the internal energy of Eq. (1) to its
corresponding free energy is to insert all the energies provided by Equations
(2), (3) and (4) into Monte-Carlo (MC) simulations. Here and technically, we employ a Metropolis Monte
Carlo (MC) algorithm implemented within the NPT or NVT (strain clamped condition) ensemble, which
samples the system's accessible thermodynamic states at a fixed temperature
($T$). The instantaneous state of the system is defined by the seven
continuous degrees of freedom (DOFs) of the Landau-type model: the
polarization-related parameter $u_{shift}$ and the six homogeneous strain
components ($\eta_{H,1,1}$, $\eta_{H,2,2}$, $\eta_{H,3,3}$, $\eta_{H,1,2}$,
$\eta_{H,1,3}$, and $\eta_{H,2,3}$).  All these
variables fluctuate during the NPT simulations while only $u_{shift}$ varies while the strain components are frozen in the NVT simulations.
The governing equation  is the total internal energy functional, $E_{tot}$, as defined by
the sum of the polar, elastic, and coupling energies in Eqs. (1-4). 
The simulation evolves by iteratively proposing a trial move,
which consists of applying a small, random displacement to a single, randomly
selected DOF. The resulting change in total energy, $\Delta E_{tot} =
E_{tot}(\text{new}) - E_{tot}(\text{old})$, is calculated. This move is then
stochastically accepted or rejected based on the criterion
that a move is always accepted if $\Delta E_{tot} \le
0$, and accepted with a probability $p = \exp(-\Delta E_{tot} / k_B T)$ if
$\Delta E_{tot} > 0$. This process ensures the system correctly samples the
Boltzmann distribution. During the MC simulations, we gradually raise the temperature of the
system from 20K to 1500K. For each temperature, we attempt to move all the
variables up to 2,560,000 times. 

Throughout the simulations, we track the
current values of all seven DOFs. The thermal averages of these
observables are then used to derive the temperature-dependent
properties shown in Fig. 9.  Obtaining the strain elements results in
the determination of the $a$ and $c$ lattice parameters, and hence the $c/a$
axial ratio, when recalling that the a$_{ref}$ and c$_{ref}$ of Table 1
correspond to the zeroes of strain for $\eta_{H,1,1}$, $\eta_{H,2,2}$ and
$\eta_{H,3,3}$.  We also note that the polarization $P_z$
is not a direct DOF but a derived observable, calculated at each step
from $u_{shift}$ and $a$ (where $a =
a_{ref}(1+\eta_{H,1,1})$) using the relation $P_{z} = 4Z^{*}u_{shift} /
(\sqrt{3}a^{2})$ with $Z^{*}$=3.2e. The average polarization $\langle P_z \rangle$ and lattice
parameters ($a$ and $c$) are computed from the thermal averages of these
tracked values. Finally, the dielectric susceptibility $\chi$ is computed from
the statistical polarization fluctuations via the
fluctuation-dissipation theorem: $\chi = \frac{V}{\varepsilon_0k_B T} \left( \langle P_z^2
\rangle - \langle P_z \rangle^2 \right)$, where $V$ is
the volume, T is the temperature, $k_B$ the Boltzmann constant and  $< >$
denotes averaging over the MC sweeps (see Ref. \onlinecite{Inna-Resta} and references
therein for that equality).

\begin{figure*} \includegraphics[width=0.9\columnwidth]{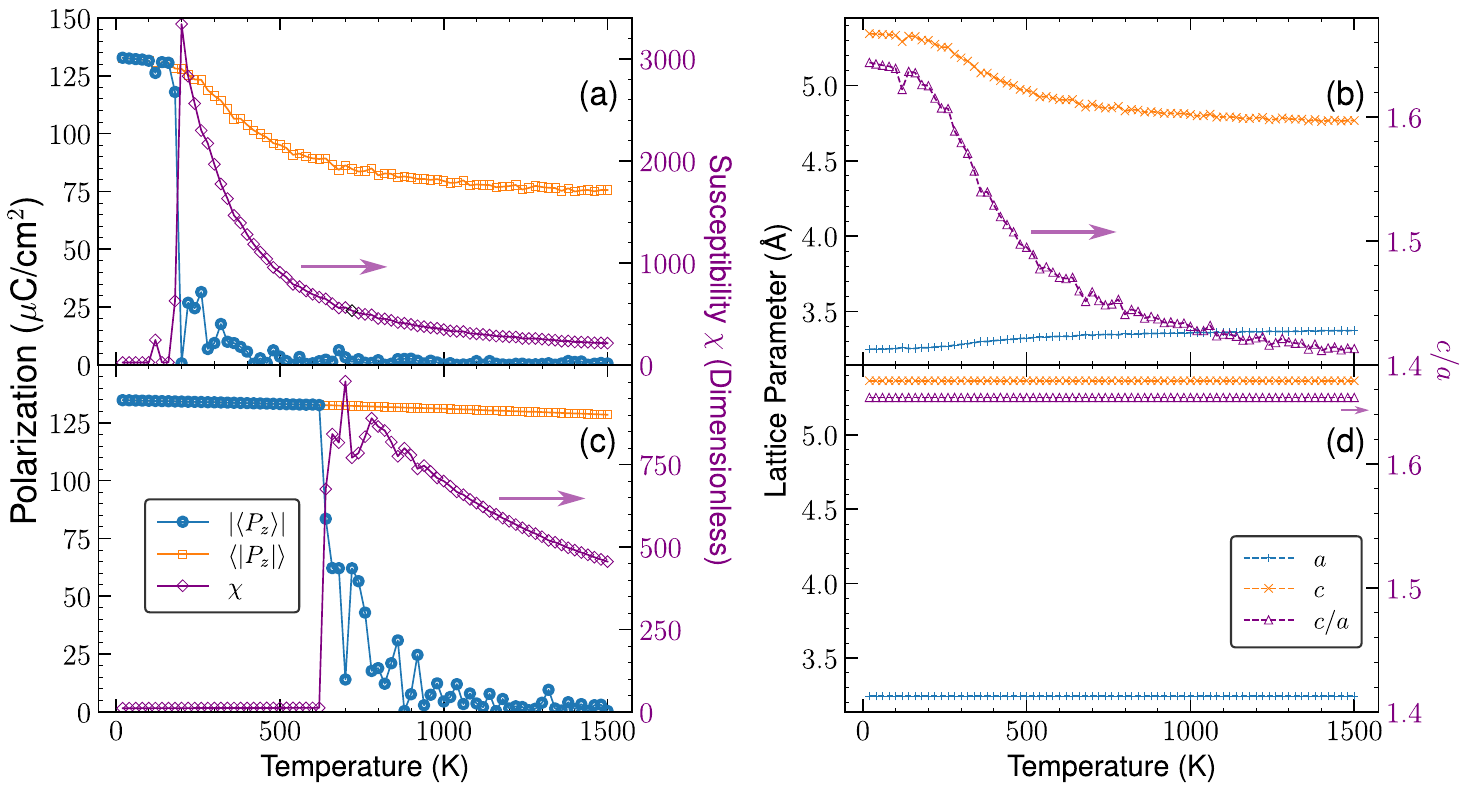}
\caption{Properties of Al$_{0.8}$Sc$_{0.2}$N as a function of temperature, as
obtained by the model of Eqs (1)-(4): (a) magnitude of the polarization,
$|<Pz>|$, average of the absolute value of the polarization,  $<|Pz|>$,
and dielectric susceptibility, (b) lattice parameters and axial ratio,  when all
degrees of freedom of this model are allowed to relax. Panels (c) and (d) 
are the same as Panels (a) and (b), respectively, but for clamped
conditions }\label{fig7} \end{figure*}

The results for Al$_{0.8}$Sc$_{0.2}$N are shown
in Figs. 9a, 9b and 9c, when allowing all the degrees of freedom to relax. These MC simulations do find a ground state that is of hexagonal symmetry since
we numerically find that all shear strains are zero while
$\eta_{H,1,1}$=$\eta_{H,2,2}$ differs from $\eta_{H,3,3}$. Figures 9a and 9b
show that, for this ground state, the resulting $c/a$ is close to the wurtzite
ideal value of 1.633 and the magnitude of the polarization is 133\,$\mu
C$/cm$^{2}$ at 20\,K, which agrees with those observed in Refs. \cite{Zhu,Drury}
yielding 120-140 $\mu C$/cm$^{2}$. Note that the $u_{shift}$ and $c/a$ of this
wurtzite ground state, which are equal to 0.118 and 1.644, respectively, are
slightly larger than the ones predicted by DFT-VCA and given in Table 2 because
the model of Eq. (1) does not reproduce at 100\% all the quantitative DFT
results -- which can also be seen in Fig. 4. Slightly varying some parameters
of Eqs (2)-(4) can result in a perfect agreement for the wurtzite ground state
between the MC results based on Eqs (1)-(4) at low temperatures and the DFT-VCA
predictions, but this article is rather aimed at qualitatively understanding
ferroelectric nitrides with a good enough accuracy and the overall agreement
between this model and the DFT-VCA results shown in Figs 4 and 5 is already
satisfactory for this purpose.

The major problems evidenced in Figs 9a and 9b are rather that the MC computations
predict that (1) the magnitude of the polarization of Al$_{0.8}$Sc$_{0.2}$N
begins to strongly deviate from $\simeq$ 130\,$\mu C$/cm$^{2}$ starting at 180K
(at which $|<Pz>|$ begins to
differs from the average of the absolute value of the polarization  $<|Pz|>$)
and is even less than 50 C/cm$^{2}$ at 300K. These results severely contrast
with measurements conducted from about $\simeq$ 293K up to $\simeq$ 523K in
Al$_{0.8}$Sc$_{0.2}$N and that still see a polarization of about 120 $\mu
C$/cm$^{2}$ at any of these temperatures \cite{Drury}; (2) the  dielectric
 susceptibility peaks at about 200K, while Ref. \cite{Islam} observes peaks in the
real part of the relative permittivity for temperatures between $\simeq$ 850K
and 1000K, when the frequency of the applied field is ranging between 1kHz and
100kHz; (3) the $c/a$ axial ratio rapidly decreases with temperature, reaching
values close to 1.4 at about  1500K, as a result of the fast
temperature-driven decrease of the $c$ lattice constant accompanied by a
concomitant increase of the $a$ lattice parameter. This is sharp contrast with
the experimental finding that Al$_{1-x}$Sc$_{x}$N disordered solid solutions,
with $x$ close to 0.2-0.3, basically keep their same axial ratio of $\simeq$
1.6 from room temperature up to the highest investigated temperature of
$\simeq$ 1400K \cite{Islam}, as a result of the fact that both $a$ and $c$ only
very weakly increase with temperature since the thermal expansion of
nitride ferroelectrics has been observed \cite{Lu} to be as weak as 4.3-4.7
$\times10^{-6}$ K$^{-1}$ (such expansion yields an increase of the $a$ and $c$
lattice parameters as small as 0.02-0.03 \AA~when heating AlN from 0K to
1400K \cite{Figge}).

One can envision two possibilities to explain the discrepancy between our
simulations and measurements.  One possibility is that the dramatic change of
lattice parameters, of the order of several tenth of \AA, shown in Fig. 9b
would result in a large induced stress that will break the sample.  Another
possibility is that  Eqs (1)-(4), when put into Monte-Carlo simulations, are
unable to correctly describe the thermal expansion of Al$_{1-x}$Sc$_{x}$N
compounds, which is not so surprising when realizing that a correct description
of such expansion is known to require the participation of all phonon modes
while only one optical zone-center mode and acoustic modes near the zone-center
are taken into account in these equations via the degree of freedoms
$u_{shift}$ and the homogeneous strain elements. 

Based on these two possibilities, we thus decided to pursue another
alternative, that is to freeze at any temperature
$\eta_{H,1,1}$=$\eta_{H,2,2}$ and  $\eta_{H,3,3}$, and therefore the $a$ and
$c$ lattice constants and the $c/a$ axial ratio, as equal to their
low-temperature equilibrium values given by the model, as shown in Fig. 9d (we
therefore neglected here thermal expansion because it is known to be
negligible, of the order of hundredth of \AA~ up to 1400K, with respect to the
change of the lattice parameters wrongly predicted in Fig. 9b and that is of the
order of several tenth of \AA). The resulting polarization of
Al$_{0.8}$Sc$_{0.2}$N  is depicted in Fig. 9c, and is now basically
independent  of temperature with a value close to 130\,$\mu C$/cm$^{2}$, up to
a temperature of $\simeq$ 640 K
(temperature at which $|<Pz>|$ starts to deviate from  $<|Pz|>$), which now
agrees with the measurement of  Ref. \cite{Drury}.  Moreover and as shown in
Fig. 9c, the dielectric  susceptibility now peaks at around 700K, which is more
consistent with the observation of Ref. \cite{Islam}. Moreover, Figs 9c and  9d
carry an important message, namely that an  axial ratio with a value typically
associated with a wurtzite structure (that is of the order of 1.5-1.7) does not
necessarily mean a large polarization. As a matter of fact, our simulations
under clamped conditions provide, e.g., for the same axial ratio of around 1.64
a large value of the polarization of 130\,$\mu C$/cm$^{2}$  up to 640K  
but also an intermediate value of
50\,$\mu C$/cm$^{2}$  for T=750K   and a vanishing value for $|<Pz>|$ near 1500K (at which a real
sample would likely have melted). In other words, Al$_{1-x}$Sc$_{x}$N compounds
can exhibit a decoupling between polarization and axial ratio at high enough
temperature. This is in contrast with perovskite oxides for which any upward
deviation of 1 of the axial ratio automatically implies the existence of a
polarization.  In other words, ferroelectric nitrides can behave differently than perovskite oxides in terms of strain-dipole coupling, and
it is questionable to assume that a ferroelectric nitride has a large polarization at finite temperature if its axial ratio is
close to the one of the ideal wurtzite structure. 

Let us now try to understand the dramatic differences shown in Figs. 9 between
the results of a free relaxation of all parameters of the model of Eqs (1-4)
and  those for clamped conditions. For that, Figures 10 displays $u_{shift}$,
axial ratio and resulting internal energy that Al$_0.8$Sc$_{0.2}$N possesses
during the MC sweeps at four different temperatures, namely 20K, 300K, 900K and
1500K, when all degrees of freedom can relax. Figures 11 display similar
results but when the lattice parameters are frozen.  Figures 10
indicate that the system is basically stuck in a WZ minimum at low temperatures (e.g., 20K)
before being able, at, e.g., 300K,  to spend time in the two opposite WZ minima
(one with a negative $ u_{shift}$ of about  -0.12 and the other one with a
positive $u_{shift}$ $\simeq$ +0.12), as well as in intermediate states
including around the non-polar H5 state  for which $c/a$ is close to 1.2.
Consequently, the polarization at 300K is much smaller than that at 20K. When the
temperature further increases to, e.g., 900K and 1500K, much more states with
$u_{shift}$ varying between -0.15 and +0.15 and  $c/a$ varying between $\simeq$
1.0 and 1.8,  are accessed and become occupied more frequently  because of their close energy. As a result,
$|<Pz>|$ vanishes and $c/a$ converges towards 1.4  at these high temperatures.
 Overall, the relaxation of mechanical degrees of freedom in the NPT ensemble authorizes the system to access low energy lying states with vastly different magnitude and sign of the polarization (see Figure 10). This facilitates the transition from positive to negative polar states even at moderate temperature and therefore lowers the transition temperature. In contrast, clamped conditions restrict the accessible potential energy surface to a line with high-energy intermediate states, preventing an easy transition path from positively polarized to negatively polarized states and consequently maintaining a large polarization at higher temperatures.


On the other hand and while the results of the trajectories are similar to the
case of full relaxation at low temperature, the clamped case displays a
different scenario for other temperatures (note that freezing the strain degrees of freedom within the NVT simulations implies that the elastic energy does not vary 
but it is finite rather than null, and the other two energies ($E_{polar}$ and $E_{int}$) naturally change when $u_{shift}$ varies during these simulations). Basically, for clamped conditions, Al$_{0.8}$Sc$_{0.2}$N
can only access the WZ minima up to 1000K, which explains why $<|Pz|>$)
is basically independent of the temperature from 20K to 1000K unlike in the
case of full relaxation (compare Fig. 9c with Fig. 9a). As the temperature
increases to 900K, the WZ minimum that was initially
unoccupied at 20K begins to be progressively filled more and more in detriment
of the WZ minimum that was fully occupied at low temperatures -- which also
results in $|<Pz>|$  decreasing and moving further away from  $<|Pz|>$. In
other words, the clamped case can be considered as an example of a strong
ordered-disordered type of transition (see Ref. \cite{Inna-Jirka} and
references therein), unlike when all the degrees of freedom can relax.  One
experimental way to check such prediction will be to observe central modes,
rather than soft modes, for ferroelectric nitrides, since central modes are
typical signatures of ordered-disordered type of transition \cite{Inna-Jirka}.
One can also note from Fig. 11 that, at very high temperatures such as 1500K, states other than the two WZ minima
(including some with zero polarization) can be reached.

By comparing Figures 10 and 11, we further observe the critical role of strain in
influencing the temperature-dependent behavior of polarization. When strain is
permitted to vary, the energy barrier between the symmetric states $u_{shift}$
and $-u_{shift}$ is significantly reduced at 300 K (Fig. 10), as compared to the
fixed-strain scenario (Fig. 11). This reduction suggests that polarization reversal can be facilitated by strain variations,
if the materials allow large change of strains to occur, which in turn should decrease the coercive field, and therefore lead to lower operating voltage of the AlScN memory elements in devices.
Consequently, such dynamic strain pathways should enable polarization reversal at lower temperatures and potentially enhance the material's
responsiveness and performance.

\begin{figure}
    	\centering
    	\includegraphics[scale=0.6]{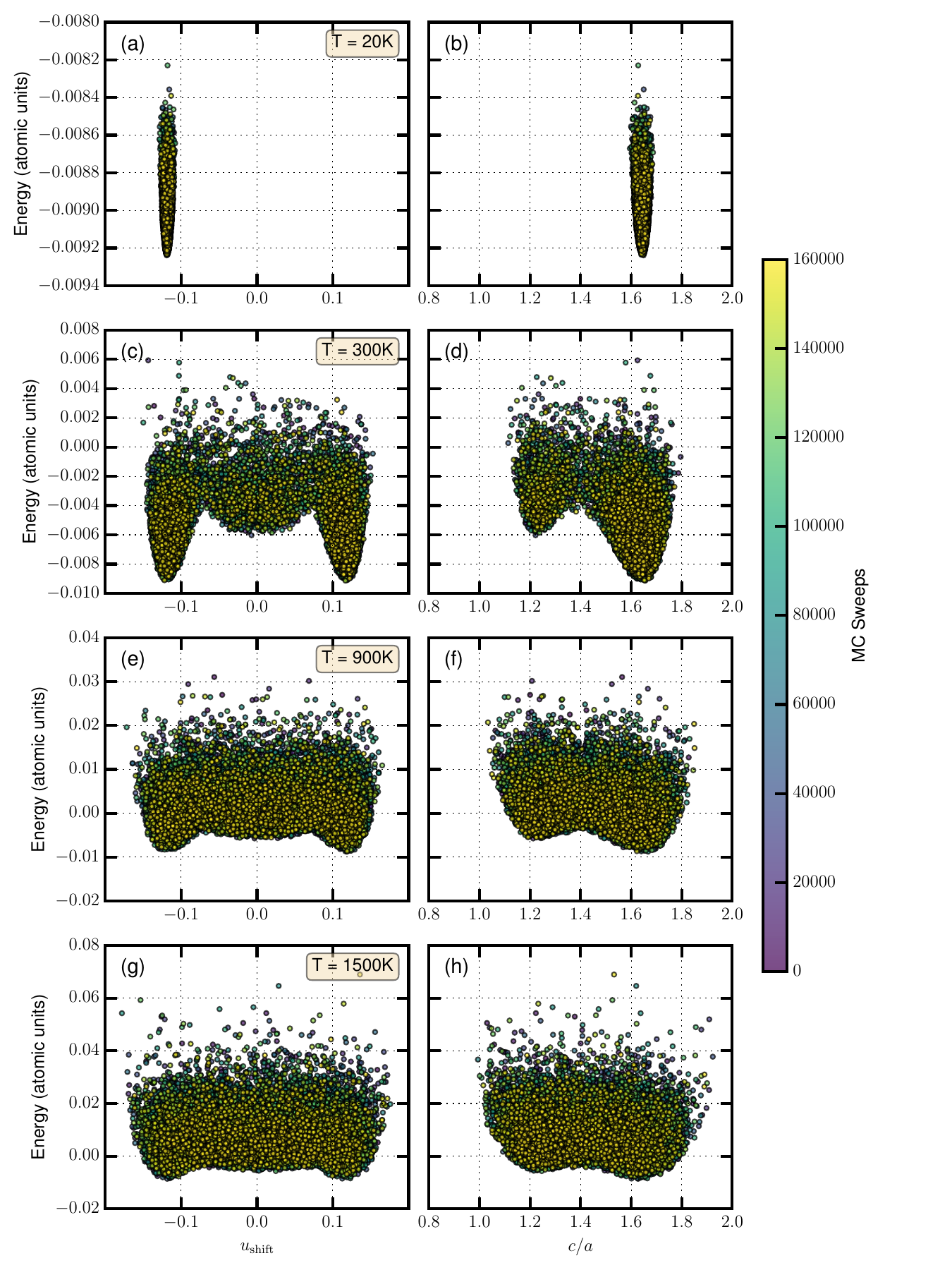}
    	\caption{Energy  landscape of Al$_{0.8}$Sc$_{0.2}$N for different temperatures, as obtained by the model of Eqs (1)-(4) and when all degrees of freedom of this model are allowed to relax: (a)
	Energy versus $u_{shift}$,  and (b) Energy versus axial ratio  at 20K, 
	 Panels (c)-(d), Panels (e)-(f) and Panels (g)-(h) are similar to Panels (a)-(b) but for 300K, 900K and 1500K, respectively.  \label{fig8}}
	\end{figure}

\begin{figure}
    	\centering
    	\includegraphics[scale=0.6]{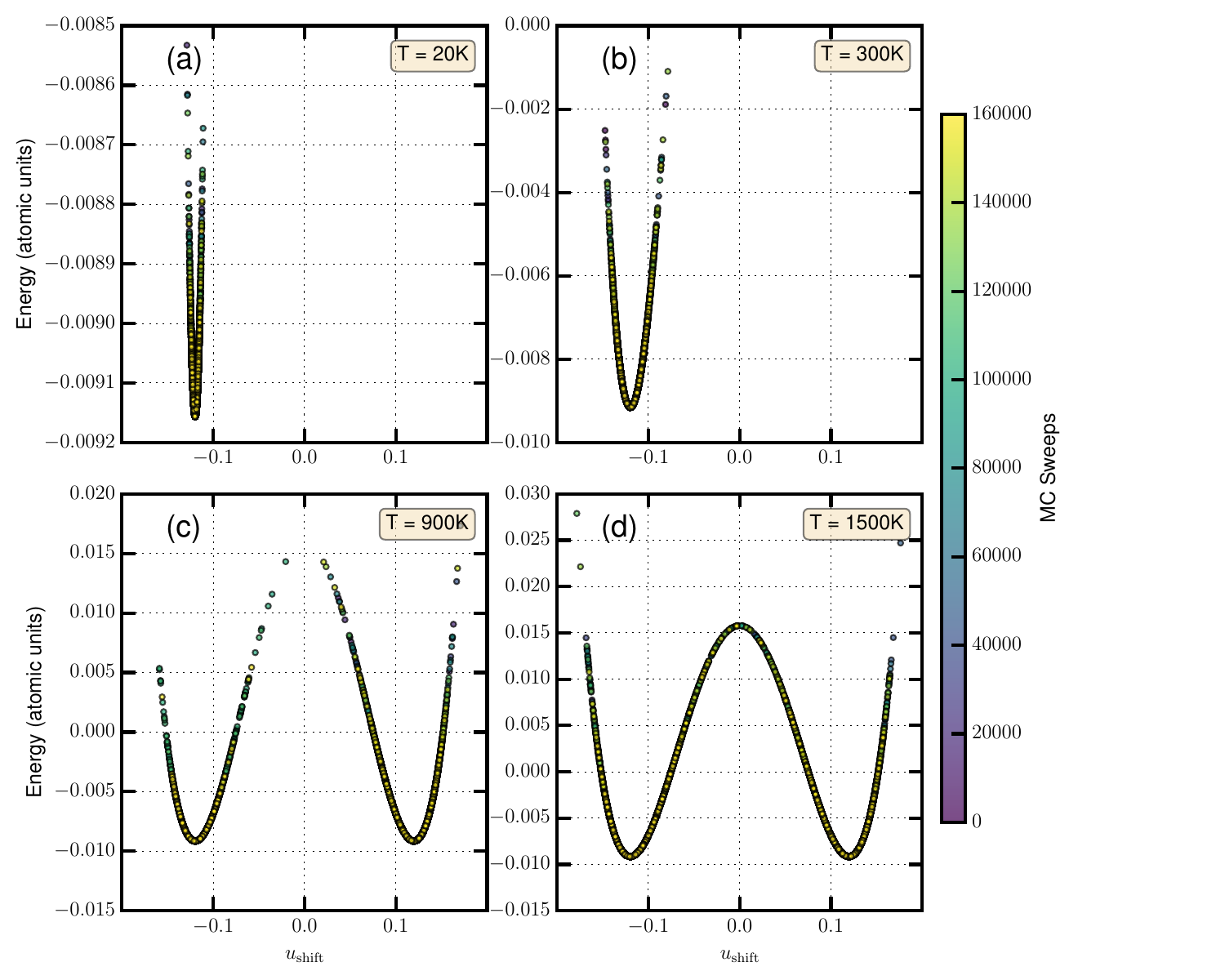}
    	\caption{Energy  landscape of Al$_{0.8}$Sc$_{0.2}$N for different temperatures, as obtained by the model of Eqs (1)-(4) but
for clamped conditions: 
	Energy versus $u_{shift}$ at 20K, 300K, 900K and 1500K for Panels (a), (b), (c) and (d), respectively.
	 \label{fig9}}
	\end{figure}

\section{Summary}

In summary, we have conducted first-principles computations and developed a
phenomenological model, with parameters extracted from {\it ab-initio}, to gain
insight into ferroelectric nitrides, in general, and (Al,Sc)N solid solutions,
in particular. This study has many take-away messages. For instance, (1) that
using the virtual crystal alloy approximation within density-functional theory
can yield structural and polar properties in excellent agreement with
experiments but also suggests that grown Al$_{1-x}$Sc$_{x}$N samples with Sc
compositions ranging between $\simeq$ 25\% and $\simeq$ 45\% may possess
regions of H5 and/or rocksalt phases in addition to WZ; (2)  a Landau-like
model can successfully describe energetics and structural properties of these
compounds at 0K both for the 5-times-coordinated H5 and 4-times-coordinated WZ
phases but also for intermediate states, when extracting its parameters from
DFT calculations; (3)  such model has to contain three main types of energy to
be able to reproduce DFT data, namely a polar energy, an elastic energy and a
third one describing the interactions between polarization and strains; (4) the
elastic energy differs from that typically employed for perovskite oxides, by
the need to possess high-order terms in strains but also by the presence of
linear terms in strains (due to the hexagonal symmetry and the chosen reference
state) \cite{Zhong,Landau}, and is the one solely responsible for the H5
minimum -- therefore making this latter of ferroelastic origin;  (5) the energy
describing the coupling between  polarization and strains is crucial to make
the WZ minimum appearing and is the only one out of the three that is
attractive in nature, unlike in ferroelectric perovskite oxides (for which the
polar energy is also attractive). One can therefore think of the WZ state as
being a trigger-type phase;  (6) this coupling energy is about three times
larger for the WZ equilibrium state of  Al$_{0.8}$Sc$_{0.2}$N than for the
ferroelectric tetragonal ground state of PbTiO$_3$ perovksite oxides; (7) it is
the compositional change of one parameter appearing in the polar energy and of
another parameter involved in the elastic energy that mostly explains why
adding Sc makes the hexagonal minimal energy going from WZ to H5; (8) the
Landau-like model can also be employed to easily search and discover low-energy
states, and to find minimal energy paths; (9) employing this latter model in
techniques such as Monte-Carlo (or Molecular Dynamics) can reproduce the
observed behavior of the polarization with temperature, but once imposing the
lattice parameters to be frozen or not to vary too much (as in experiments).
This fact is to be  remembered if one desires to extend such model to atomistic
effective Hamiltonians that will incorporate short-range and long-range polar
and elastic interactions (i.e., between different cells or within the different
ions belonging to the same unit cell). These hypothetical effective
Hamiltonians would then be able to tackle problems that are not feasible with
the present  Landau-type model, such as the switching of the polarization and
its mechanisms -- which constitute a timely and important topic
\cite{Lee1,Konishi,Liu,Krishnamoorthy}; (10) the use of this model under
clamped conditions suggests another astonishing result, namely that the strains
and overall polarization can be decoupled at high temperature, because the
paraelectric-to-polar transition is of the ordered-disordered type. The concept
of soft mode is therefore likely not applicable to (Al,Sc)N, unlike that of
central mode describing jumps between potentials,  and measuring an axial ratio
of about 1.6 does not necessarily mean that the system possesses a large
polarization at high temperature -- unlike what is commonly currently believed
\cite{Islam}.  Let us also end this article by emphasizing that Equations (1-4)
should also be relevant to ferroelectric II-VI oxides such as (Zn,Mg)O
compounds \cite{Ferri} in their hexagonal form since pure ZnO is known to adopt
a wurtzite state and that a metastable H5 phase also occurs in MgO
\cite{Limpijumnong}. We thus hope that this article is of relevance  to a broad
community interested in, e.g., ferroelectrics, nitrides, semiconductors,
modelisation, neuromorphic computing, and will encourage experimentalists to
check our predictions.

The authors thank  the U.S. Department of Energy (DOE), Office of Science,
Basic Energy Sciences (BES) under Award No. DE-SC0025479. 
AZ and KY also acknowledge DOE Office of Science (SC), Basic Energy Sciences (BES), Materials Chemistry program. 
This work was co-authored by National Renewable Energy Laboratory for the US Department of Energy (DOE) under Contract No. DE-AC36-08GO28308. 
The views expressed in the article do not necessarily represent the views of the DOE or the US Government. This work was
conducted fully or in part with the support of the Arkansas High Performance
Computing Center which is funded through multiple National Science Foundation
grants and the Arkansas Economic Development Commission.


\begin{thebibliography}{99}

\bibitem{Farrer}  N. Farrer and L. Bellaiche. Properties of hexagonal ScN
versus wurtzite GaN and InN. Physical Review B {\bf 66}, 201203(R) (2002).

\bibitem{Nakamura} S. Nakamura. InGaN-based blue/green LEDs and laser diodes.
Advanced Mater {\bf 8}, 689 (1996).

\bibitem{Edgar} Properties, Processing and Applications of Gallium Nitride and
Related Semiconductors, edited by J. H. Edgar, S. Strite, I. Akasaki, H. Amano,
and C. Wetzel, Emis Datareviews Series No. 23 (Inspec, London, 1999).

\bibitem{Wright} A. F. Wright and J. S. Nelson. Consistent structural
properties for AlN, GaN, and InN, Physical Review B {\bf 51}, 7866 (1995).

\bibitem{Ohba} N. Ohba, K. Miwa, N. Nagasako, and A. Fukumoto. First-principles
study on structural, dielectric, and dynamical properties for three BN
polytypes. Physical Review B 63, 115207 (2001).

\bibitem{Ranjan} V. Ranjan, L. Bellaiche and E.J. Walter. Strained hexagonal
ScN: A material with unusual structural and optical properties. Physical Review
Letters {\bf 90}, 257602 (2003).

\bibitem{Ranjan2} V. Ranjan, S. Bin-Omran, L. Bellaiche and A. Alsaad.
Isostructural phase transitions in GaN/ScN and InN/ScN superlattices. Physical
Review B {\bf 71}, 195302 (2005).

\bibitem{Ranjan3} V. Ranjan, S. Bin-Omran, D Sichuga, R. S. Nichols. Properties
of GaN/ScN and InN/ScN superlattices from first principles. Physical Review B
{\bf 72}, 085315 (2005).

\bibitem{Daoust} P. Daoust, P. Desjardins, R. A. Masut, V. Gosselin, and M.
Cot\'e, Ab-initio piezoelectric properties of Al$_{0.5}$Sc$_{0.5}$N: Impact of
alloy configuration on the d$_{33,f}$ piezoelectric strain coefficient,
Physical Review Mater {\bf 1}, 055402 (2017).

\bibitem{Tasnadi} F. Tasnadi, B. Alling, C. Hoglund, G. Wingqvist, J. Birch, L.
Hultman, and I. A. Abrikosov. Origin of the Anomalous Piezoelectric Response in
Wurtzite Sc$_{x}$Al$_{1-x}$N Alloys. Physical Review Letters {\bf 104}, 137601
(2010).


\bibitem{Noor} M. Noor-A-Alam, O. Z. Olszewski, and M. Nolan, Ferroelectricity
and large piezoelectric response of AlN/ScN superlattice, ACS Applied Materials
and Interfaces {\bf} 11, 20482 (2019).


\bibitem{Zhijun} Z. Jiang, C. Paillard, D. Vanderbilt, H. Xiang and L.
Bellaiche. Designing multifunctionality via assembling dissimilar materials:
Epitaxial AlN/ScN superlattices. Physical Review Letters {\bf 123}, 096801
(2019).


\bibitem{Zhijun2} Z. Jiang, B. Xu, H. Xiang, and L. Bellaiche. Ultrahigh energy
storage density in epitaxial AlN/ScN superlattices. Physical Review Materials
{\bf 5}, L072401 (2021).


\bibitem{Akiyama2} M. Akiyama, K. Kano, and A. Teshigahara, Influence of growth
temperature and scandium concentration on piezoelectric response of scandium
aluminum nitride alloy thin films. Applied Physics Letters {\bf 95}, 162107
(2009).

\bibitem{Fichtner} S. Fichtner, N. Wolff, F. Lofink, L. Kienle, and B. Wagner.
AlScN: A III-V semiconductor based ferroelectric, Journal Applied Physics {\bf
125}, 114103 (2019).


\bibitem{Wolff} N. Wolff, G. Schonweger, I. Streicher, M. R. Islam, N. Braun,
P. Stranak, L. Kirste, M. Prescher, A. Lotnyk, H. Kohlstedt, S. Leone, L.
Kienle, and S. Fichtner. Demonstration and STEM Analysis of Ferroelectric
Switching in MOCVD-Grown Single Crystalline Al0.$_{85}$Sc$_{0.15}$N. Advanced
Physics Research {\bf 3}, 2300113 (2024).

 \bibitem{Islam} Md. R. Islam {\it et al.}. On the exceptional temperature
stability of ferroelectric Al$_{1-x}$Sc$_{x}$N thin films. Applied Physics
Letters {\bf 118}, 232905 (2021).
 
 
\bibitem{Guido} R. Guido, P. D. Lomenzo, M. R. Islam, N. Wolff, M. Gremmel, G.
Schonweger, H. Kohlstedt, L. Kienle, T. Mikolajick, S. Fichtner, and U.
Schroeder. Thermal Stability of the Ferroelectric Properties in 100 nm-Thick
Al$_{0.72}$Sc$_{0.28}$N. ACS Applied Materials and Interfaces {\bf 15},
7030 (2023).


\bibitem{Zhu} W. Zhu, J. Hayden, F. He, J.-I. Yang; P. Tipsawat, M. D. Hossain,
J.-P. Maria, S. Trolier-McKinstry. Strongly temperature dependent ferroelectric
switching in AlN, Al$_{1-x}$Sc$_{x}$N, and Al$_{1-x}$B$_{x}$N thin films.
Applied Physics Letters {\bf 119}, 062901 (2021).



\bibitem{Yazawa1} K. Yazawa, J. Hayden, J.-P. Maria, W. Zhu, S.
Trolier-McKinstry, A. Zakutayev, and G. L. Brennecka. Anomalously abrupt
switching of wurtzite-structured ferroelectrics: simultaneous non-linear
nucleation and growth model. Materials Horizon {\bf 10}, 2936 (2023).

\bibitem{Yazawa2} K. Yazawa, D. Drury, A. Zakutayev, and G. L. Brennecka.
Reduced coercive field in epitaxial thin film of ferroelectric wurtzite
Al$_{0.7}$Sc$_{0.3}$N. Applied Physics Letters {\bf 118}, 162903 (2021).

\bibitem{Lee1} C.-W. Lee, K. Yazawa, A. Zakutayev, G. L. Brennecka, and P.
Gorai. Switching it up: New mechanisms revealed in wurtzite-type
ferroelectrics. Science Advances {\bf 10} eadl0848 (2024).


\bibitem{Lee2} C.-W. Lee, N. Ud Din, K. Yazawa, G. L. Brennecka, A. Zakutayev,
and P. Gorai. Emerging Materials and Design Principles for Wurtzite-type
Ferroelectrics. ChemRxiv:2023-hf-60w (2023).

\bibitem{Calderon} S. Calderon, J. Hayden, S. M. Baksa, W. Tzou, S.
Trolier-McKinstry, I. Dabo, J.-P. Maria, and E. C. Dickey. Atomic-scale
polarization switching in wurtzite ferroelectrics. Science  {\bf 380}, 1034
(2023).


\bibitem{Drury} D. Drury , K. Yazawa, A. Zakutayev, B. Hanrahan  and G.
Brennecka. High-Temperature Ferroelectric Behavior of Al$_{0.7}$Sc$_{0.3}$N.
Micromachines {\bf 13}, 887 (2022).

\bibitem{Drury2} D. Drury, K. Yazawa, G. Brennecka, B. Hanrahan. 
Ferroelectric epitaxial Al(Sc/B)N/Mo/SiC heterostructures for high operating temperature devices.
Journal of Applied Physics {\bf 137}, 204102 (2025).


\bibitem{Kei-Andriy} K. Yazawa, A. Zakutayev and G. L. Brennecka. A
Landau-Devonshire analysis of strain effects on ferroelectric
Al$_{1-x}$Sc$_{x}$N.  Applied Physics Letters {\bf 121}, 042902 (2022).

\bibitem{Satoh} S. Satoh, K. Ohtaka, T. Shimatsu and S. Tanak. Crystal structure
deformation and phase transition of AlScN thin films in whole Sc concentration
range. Journal of Applied Physics {\bf 132}, 025103 (2022).


\bibitem{Giannozzi2009} P. Giannozzi {\it et al.}. Quantum Espresso: a modular
and open-source software project for quantum simulations of materials.  Journal
of Physics: Condensed Matter {\bf 21}, 395502 (2009).

\bibitem{Perdew1996} J.P. Perdew,  K. Burke and M. Ernzerhof. Generalized
Gradient Approximation Made Simple. Physical Review Letters {\bf 77}, 3865
(1996).

\bibitem{Hamann2013} D.R. Hamann. Optimized norm-conserving Vanderbilt
pseudopotentials. Physical Review B {\bf 88}, 085117 (2013).

\bibitem{Monkhorst1976} H.J. Monkhorst and J.D. Pack.  Special points for
Brillouin-zone integrations. Physical Review B {\bf 13}, 5188  (1976).


\bibitem{Bellaiche2000} L. Bellaiche and D. Vanderbilt. Virtual crystal
approximation revisited: Application to dielectric and piezoelectric properties
of perovskites. Physical Review B {\bf 61}, 7877 (2000).



\bibitem{Akiyama} M. Akiyama, T. Kamohara, K. Kano, A. Teshigahara, Y. Takeuchi
and N. Kawahara, Enhancement of Piezoelectric Response in Scandium Aluminum
Nitride Alloy Thin Films Prepared by Dual Reactive Cosputtering. Advanced
Materials {\bf 21}, 593 (2009).   

\bibitem{Andryinew} K. Yazawa, J. S. Mangum, P. Gorai, G. L. Brennecka and  A. Zakutayev Local chemical origin of ferroelectric behavior in wurtzite nitrides. Journal of Materials Chemistry C {bf 10}, 17557 (2022).

 \bibitem{Landau} G. Shirane and F. Jona. Ferroelectric Crystals (Macmillan,
New York, 1962).
 
 \bibitem{Zunger1990} A. Zunger, S. H. Wei.  L. G. Ferreira and J. E. Bernard.
Special quasirandom structure. Physical Review Letters {\bf 65}, 353 (1990).

\bibitem{vandeWalle2013} A. van de Walle, P. Tiwary, M. de Jong, D.L. Olmsted,
M. Asta, A. Dick, D. Shin, Y. Wang, L.-Q. Chen and Z.-K. Liu. Efficient
stochastic generation of special quasirandom structures. Calphad {\bf 42}, 13
(2013).

\bibitem{Kresse1994} G Kresse and J Hafner. Ab initio molecular-dynamics
simulation of the liquid-metal-amorphous-semiconductor transition in germanium.
Physical Review B {\bf 49}, 14251 (1994).

\bibitem{Kresse1996} G Kresse and J. Furthmuller. Efficiency of ab-initio total
energy calculations for metals and semiconductors using a plane-wave basis set.
Computational Materials Science {\bf 6}, 15 (1996).

\bibitem{Kresse1999} G Kresse and D Joubert. From ultrasoft pseudopotentials to
the projector augmented-wave method. Physical Review B {\bf 59}, 1758 (1999).

\bibitem{Blochl1994} P. E Blochl. Projector augmented-wave method. Physical
Review B {\bf 50}, 17953 (1994).

\bibitem{Domenic2}  R.D. King-Smith and D. Vanderbilt. Theory of Polarization of
Crystalline Solids.  Physical Review B {\bf 47}, 1651 (1993).

\bibitem{Resta} R. Resta. Macroscopic polarization in crystalline dielectrics: the geometric phase approach.
 Reviews of Modern Physics {\bf 66}, 899 (1994).

\bibitem{ionicradius} Please see $https://mrlweb.mrl.ucsb.edu/~seshadri/Periodic/index.html$


\bibitem{Zhong} W. Zhong, D. Vanderbilt, and K. M. Rabe. Phase Transitions in
BaTiO$_3$ from First Principles. Physical Review Letters {\bf 73}, 1861 (1994);
W. Zhong, D. Vanderbilt, and K. M. Rabe. First-principles theory of
ferroelectric phase transitions for perovskites: The case of BaTiO$_3$ Physical
Review B {\bf 52}, 6301 (1995).

\bibitem{triggered1} J. Holakovsky. A new type of ferroelectric phase
transition. Physics Status Solidi B {\bf 56}, 615 (1973).

\bibitem{triggered2} I. A. Kornev and L. Bellaiche. Nature of the ferroelectric
phase transition in multiferroic BiFeO$_3$ from first principles. Physical
Review B {\bf 79}, 100105(R) (2009).


\bibitem{15} D. Fausti, R. I. Tobey, N. Dean, S. Kaiser,  A. Dienst,  M. C.
Hoffmann,    S. Pyon, T. Takayama,  H. Takagi, and A. Cavalleri Light-Induced
Superconductivity in a Stripe-Ordered Cuprate. Science {\bf 331}, 189  (2011).

\bibitem{16} M. Rini, R. Tobey, N. Dean, J. Itatani, Y. Tomioka, Y. Tokura, R.
W. Schoenlein, and A. Cavalleri. Control of the electronic phase of a manganite
by mode-selective vibrational excitation.  Nature {\bf 449}, 72 (2007).

\bibitem{17} L. Stojchevska,  I. Vaskivskyi,  T. Mertelj,  P. Kusar,  D.
Svetin,  S. Brazovskii,  and D. Mihailovic. Ultrafast Switching to a Stable
Hidden Quantum State in an Electronic Crystal.  Science {\bf 344}, 177 (2014).

\bibitem{18} X. Li, T. Qiu,  J. Zhang,  E. Baldini,  J. Lu,  A. M. Rappe, and
K. A. Nelson. Terahertz field-induced ferroelectricity in quantum paraelectric
SrTiO$_3$.   Science \textbf{364}, 1079 (2019).

\bibitem{19} T. F. Nova,  A. S. Disa,  M. Fechner, and A. Cavalleri. Metastable
ferroelectricity in optically strained SrTiO$_3$. Science \textbf{364}, 1075
(2019).

\bibitem{20} M. Liu, H. W. Hwang, H. Tao, A. C. Strikwerda, K. Fan, G. R.
Keiser, A. J. Sternbach, K. G. West, S. Kittiwatanakul, J. Lu, S. A. Wolf, F.
G. Omenetto, X.  Zhang, K. A. Nelson, and R. D. Averitt.
Terahertz-field-induced insulator-to-metal transition in vanadium dioxide
metamaterial.   Nature {\bf 487}, 345 (2012).


\bibitem{SergeyPMN}  S. Prosandeev, J. Grollier, D. Talbayev, B. Dkhil and L.
Bellaiche. Ultrafast Neuromorphic Dynamics Using Hidden Phases in the Prototype
of Relaxor Ferroelectrics, Physical Review Letters {\bf 126}, 027602 (2021).

\bibitem{SergeyNNO} S. Prosandeev, S. Prokhorenko, Y. Nahas, Y. Yang, C. Xu, J.
Grollier, D. Talbayev, B. Dkhil and L. Bellaiche. Hidden phases with
neuromorphic responses and highly enhanced piezoelectricity in an
antiferroelectric prototype. Physical Review B {\bf 105}, L100101 (2022).

\bibitem{SergeyBRFO} S. Prosandeev and L. Bellaiche. THz-induced activation of
hidden states in rare-earth-doped BiFeO$_3$ solid solutions, Physical Review
Materials {\bf 6}, 116201 (2022).

 \bibitem{SergeyPZTfilms}  S. Prosandeev, S. Prokhorenko, Y. Nahas, J.
Grollier, D. Talbayev, B. Dkhil and L. Bellaiche. Ultrafast Activation and
tuning of Topological Textures in Ferroelectric Nanostructures. Advanced
Electronic Materials {\bf 8}, 2200808 (2022).
 
 \bibitem{SergeyReview} S. Prosandeev, S. Prokhorenko, Y.Nahas, Y. Yang, C. Xu,
J. Grollier, D. Talbayev, B. Dkhil, and L. Bellaiche. Designing polar textures
with ultrafast neuromorphic features from atomistic simulations. Neuromorphic
Computing and Engineering {\bf 3}, 012002 (2023). 
 
 \bibitem{Boyn} S. Boyn {\it et al.}. Learning through ferroelectric domain
dynamics in solid-state synapses. Nature Communications {\bf 8}, 14736 (2017).
 
 \bibitem{21} I. Vaskivskyi, I. A. Mihailovic,  S. Brazovskii,  J. Gospodaric,
T. Mertelj,  D. Svetin,    P. Sutar, and D. Mihailovic. Fast electronic
resistance switching involving hidden charge density wave states.  Nature
Communicatiions {\bf 7}, 11442 (2016).
 
  \bibitem{Inna-Resta}  I. Ponomareva, L. Bellaiche and R. Resta. Relation
between dielectric responses and polarization's fluctuations in ferroelectric
nanostructures. Physical Review B 76, 235403 (2007).


\bibitem{Lu} Y. Lu {\it et al.}.  Elastic modulus and coefficient of thermal
expansion of piezoelectric Al$_{1-x}$Sc$_{x}$N up to x = 0.41) thin films.
Applied Physics Letters Materials {\bf  6}, 076105 (2018).

\bibitem{Figge} S. Figge, H. Kroncke, D. Hommel, and B. M. Epelbaum.
Temperature dependence of the thermal expansion of AlN. Applied Physics Letters
{\bf 94}, 101915 (2009).



 
 \bibitem{Inna-Jirka} J. Hlinka, T. Ostapchuk, D. Nuzhnyj, J. Petzelt, P.
Kuzel, C. Kadlec, P. Vanek, I. Ponomareva and L. Bellaiche. Coexistence of the
Phonon and Relaxation Soft Modes in the Terahertz Dielectric Response of
Tetragonal BaTiO3. Physical Review Letters {\bf 101}, 167402 (2008).
 
\bibitem{HeffPZT} I. A. Kornev, L. Bellaiche, P.-E. Janolin, B. Dkhil and E.
Suard. Phase diagram of Pb(Zr,Ti)O$_3$ solid solutions from first principles.
Physical Review Letters {\bf 97}, 157601 (2006).

\bibitem{HeffBST} L. Walizer, S. Lisenkov and L. Bellaiche. Finite-temperature
properties of (Ba,Sr)TiO$_3$ systems from atomistic simulations. Physical
Review B {\bf 73}, 144105 (2006).  .  \bibitem{Konishi} A. Konishi, T. Ogawa,
C. A. J. Fisher, A. Kuwabara, T. Shimizu, S. Yasui, M. Itoh, and H. Moriwake.
Mechanism of polarization switching in wurtzite-structured zinc oxide thin
films. Applied Physics Letters {\bf 109}, 102903 (2016). 

\bibitem{Liu} Z. Liu, X. Wang, X. Ma, Y. Yang, and D. Wu. Doping effects on the
ferroelectric properties of wurtzite nitrides. Applied Physics Letter  {\bf
122}, 122901 (2023). 

\bibitem{Krishnamoorthy}  A. Krishnamoorthy, S. C. Tiwari, A. Nakano, R. K.
Kalia, and P. Vashishta. Electric-field-induced crossover of polarization
reversal mechanisms in Al$_{1-x}$Sc$_x$N ferroelectrics. Nanotechnology {\bf
32}, 49LT02 (2021). 

\bibitem{Ferri} K. Ferri; S. Bachu, W. Zhu, M. Imperatore, J. Hayden, N. Alem,
N. Giebink, S. Trolier-McKinstry, J.-P. Maria. Ferroelectrics everywhere:
Ferroelectricity in magnesium substituted zinc oxide thin films. Journal of
Applied Physics {\bf 130}, 044101 (2021).

\bibitem{Limpijumnong} S. Limpijumnong and W.R.L. Lambrecht. Theoretical study
of the relative stability of wurtzite and rocksalt phases in MgO and GaN.
Physical Review B {\bf 63}, 104103 (2001).

 
 
\end{thebibliography}
\end{document}